\documentclass[12pt]{article}

\usepackage{a4wide,feynmf,amsmath,amssymb,float}
\usepackage{graphicx,psfrag,epsfig}

\newcommand{\lfrac}[2]{\textstyle\frac{#1}{#2}\displaystyle}
\newcommand{\ReTilde}{\widetilde{\rm Re}\,}
\newcommand{\ntrl}[1]{\tilde{\chi}^0_#1}
\newcommand{\chrg}[1]{\tilde{\chi}^+_#1}
\newcommand{\ps}{p\!\!\!\!\;/}

\newcommand{\bea}{\begin{eqnarray}}
\newcommand{\eea}{\end{eqnarray}}

\begin{document}

\thispagestyle{empty}
\setcounter{page}{0}
\def\thefootnote{\fnsymbol{footnote}}

{\textwidth 15cm

\begin{flushright}
KA--TP--6--2002\\
hep-ph/0203159 \\
\end{flushright}

\vspace{2cm}

\begin{center}

{\large\sc {\bf Complete one-loop corrections to the mass spectrum \\[0.3cm]
                of charginos and neutralinos in the MSSM}} 

\vspace{2cm}

{\sc T. Fritzsche$^a$} and {\sc W. Hollik$^{a,b}$ }

\vspace*{1cm}

     $^a$Institut f\"ur Theoretische Physik, Universit\"at Karlsruhe \\
     D-76128 Karlsruhe, Germany
     
\vspace*{0.4cm}
  
     $^b$Max-Planck-Institut f\"ur Physik, F\"ohringer Ring 6 \\
     D-80805 M\"unchen, Germany

\end{center}

\vspace*{2cm}

\begin{abstract}
\noindent
The mass  spectrum of the  chargino--neutralino sector in the minimal
supersymmetric standard model (MSSM) is calculated at the one-loop
level, based on the complete set of one-loop diagrams. 
On-shell renormalization conditions are applied to determine the
counterterms for the gaugino-mass-parameters $M_1, M_2$ and the
Higgsino-mass parameter $\mu$. The input is fixed in terms of three pole
masses (two charginos and one neutralino); the other pole masses receive
a shift with respect to the tree-level masses, which can amount to
several GeV.  
The detailed evaluation shows that both the fermionic/sfermionic loop
contributions and the non-(s)fermionic loop contributions are of the
same order of magnitude and are thus relevant for precision studies at
future colliders. 
\end{abstract}

}
\def\thefootnote{\arabic{footnote}}
\setcounter{footnote}{0}

\newpage

\section{Introduction}

Experiments at future high-energy colliders will be able to discover
supersymmetric particles and to investigate their properties. 
Provided their masses are not too high, a linear electron-positron
collider will be the best environment for precision studies
of supersymmetric models~\cite{TESLA}, especially of the minimal
supersymmetric standard model (MSSM). 
From precise measurements of masses, cross sections and asymmetries in
chargino and neutralino production, the fundamental parameters can be
reconstructed~\cite{reconstruct}, to shed light on the mechanism of SUSY
breaking. 

In view of the experimental prospects it is inevitable to include
higher-order terms in the calculation of the measureable quantities in
order to achieve theoretical predictions matching the experimental
accuracy. Former studies on chargino-pair production
\cite{diaz,yamada,blank-hollik} and scalar-quark decays~\cite{guasch} 
have demonstrated that Born-level predictions can be influenced
significantly by one-loop radiative corrections.
In \cite{blank-hollik,guasch}, with complete one-loop calculations
performed in  on-shell renormalization schemes, it was shown that
besides the fermion- and sfermion-loop contributions also the virtual
contributions from the supersymmetric gauge and Higgs sector are not
negligible.  

Since the masses of charginos and neutralinos are among the precision
observables with lots of information on the SUSY-breaking structure, the
relations between the particle masses and the SUSY parameters as well as
the relations between the masses themselves are important theoretical
objects for precision calculations. 
Previous studies were done in the ${\overline{\rm MS}}$ renormalization
scheme~\cite{pierce,lahanas} with running parameters.
In~\cite{eberl} an on-shell scheme has been proposed to calculate the
one-loop mass matrices $X$ and $Y$ of the chargino and neutralino
sector, which after diagonalization yield the one-loop corrected mass
eigenstates. The concrete evaluation of the mass spectrum
in~\cite{eberl} has been performed with the subset of the diagrams
involving only fermion and sfermion loops and is thus not yet complete
at the one-loop level.  

In this paper we present an on-shell calculation of the chargino and
neutralino mass spectrum of the MSSM at the one-loop level, based on the
entire set of one-loop diagrams. We specify the on-shell renormalization
scheme by treating all particle masses as pole masses, and with field
renormalization implemented in a way that allows to formulate the
renormalized 2-point vertex functions as UV-finite matrices which become
diagonal for external momenta on-shell. The masses of the two charginos
and of one neutralino are used as input to fix the MSSM parameters $\mu,
M_1, M_2$. 
Since only the gaugino-mass parameters $M_1, M_2$ and the Higgsino-mass
parameter $\mu$ can be renormalized independently in terms of three pole
masses, with all other renormalization constants fixed in the gauge and
Higgs sector, the residual eigenvalues of the tree-level mass matrices
are no longer the pole positions of the corresponding dressed
propagators; the pole masses hence receive a shift versus the tree-level
masses, which is calculable in terms of the renormalized self-energies.  
As a byproduct, we obtain all the renormalization constants required to
determine the various counterterms for the chargino--neutralino sector 
of the MSSM, being implemented in the MSSM version of 
{\tt FeynArts}~\cite{hahnschappacher} for completion at the one-loop
level. 

After explaining  the general structure of the renormalization of
parameters and fields in sections 3 and 4, we specify the on-shell
conditions in section 5 and give the explicit solutions for the
renormalization constants. 
The calculation of the predicted neutralino masses is outlined in
section 6, and a presentation and discussion of the numerical results is
given in section 7.  

\section{Notations and parameters}
\label{sec:Notations}

The bilinear part of the Lagrangian describing the 
chargino/neutralino sector of the MSSM,
\bea
\label{eqn:Lagrangian}
\mathcal{L} & = & \mathcal{L}_{\rm kin} + 
\mathcal{L}_{\rm mass} \, ,
\eea
is composed of the kinetic term
\begin{eqnarray}
\mathcal{L}_{\rm kin} & = &
  i \,\overline{\lambda}^a\,\overline{\sigma}^\mu\,
     \big(\partial_\mu \lambda\big)^a
+ i \,\overline{\lambda'}\,\overline{\sigma}^\mu\,
      \big(\partial_\mu \lambda'\big)
+ i\, \overline{\psi}_{H_1}\overline{\sigma}^\mu 
       \partial_\mu \psi_{H_1}
+ i\, \overline{\psi}_{H_2}\overline{\sigma}^\mu 
       \partial_\mu \psi_{H_2}
\label{eqn:MSSMLkin}
\end{eqnarray}
and the mass term following from the expression
\begin{eqnarray}
\mathcal{L}_{\rm mass} & = &
\sqrt{2}\,\big[
i \,H_1^\dagger\bigl(g\,\lambda^a\,T^a
+\lfrac{1}{2} \,g'\,\lambda'\bigr)\psi_{H_1}
+i\,H_2^\dagger\bigl(g\,\lambda^a\,T^a
+\lfrac{1}{2}\,g'\,\lambda'\bigr)\psi_{H_2} 
+ \mbox{h.c.}\big]
\nonumber \\ && {}
+\epsilon_{ij}\big(\mu\,\psi_{H_1}^i \psi_{H_2}^j + \mbox{h.c.}\big)
+\lfrac{1}{2}
\bigl(M_1 \lambda'\lambda' + M_2 \lambda^a\lambda^a +\mbox{h.c.}\bigr)
\label{eqn:MSSMPhotonWZ3a}
\end{eqnarray}
by substituting the vacuum configurations of the two Higgs-doublet 
fields $H_{1,2}$. 
The Lagrangian in two-component notation involves the
Weyl spinors $\lambda', \, \lambda^a \, (a=1,2,3)$ for the
gauginos and $\psi_{H_i}^{1,2}$
for the Higgsino isospin
components accompanying the components of the Higgs doublets, i.e.\
\begin{equation}
 H_i \, = \,
\left( \begin{array}{c}
 H_i^1 \\  H_i^2 
\end{array}  \right) , \quad
\psi_{H_i} \, = \,
\left(\begin{array}{c}
\psi_{H_i}^1 \\  \psi_{H_i}^2 \end{array} \right) .
\end{equation} 
Besides the gauge couplings $g$ and $g'$, the Lagrangian involves
the $\mu$ parameter,  
the soft-breaking gaugino-mass parameters $M_1$ and $M_2$, 
and the Higgs vacua $v_i$, which are related to 
$\tan\beta = v_2/v_1$ and to the $W$ mass
$M_W = g v/2$ with $v = (v_1^2 + v_2^2)^{1/2}$.

\section{Charginos}
\label{sec:Charginos}
\subsection{Lagrangian and mass eigenstates}

Introducing a compact notation by
collecting the chiral parts parts according to 
\begin{eqnarray}
\psi^R \equiv \left(\begin{array}{c}
\psi^R_1 \\ \psi^R_2
\end{array}\right)
 =\left(\begin{array}{c}
-i\,\lambda^- \\ \psi_{H_1}^2
\end{array}\right) , \quad
\psi^L \equiv \left(\begin{array}{c}
\psi^L_1 \\ \psi^L_2
\end{array}\right)
 =\left(\begin{array}{c}
-i\,\lambda^+ \\ \psi_{H_2}^1
\end{array}\right) ,
\label{eqn:RenChar1}
\end{eqnarray}
with 
$\lambda^\pm = \lfrac{1}{\sqrt{2}}\big(\lambda^1\mp i 
\lambda^2\big)$, 
leads to the conventional form of the bilinear terms of the
Lagrangian
(\ref{eqn:Lagrangian})--(\ref{eqn:MSSMPhotonWZ3a}) for the
charginos,
\begin{eqnarray}
\mathcal{L}_{\rm ch} & = 
i\bigl[
{\psi^R}^\top\sigma^\mu\partial_\mu\,\overline{\psi}^R + 
{\overline{\psi}^L}^\top\overline{\sigma}^\mu\partial_\mu\,\psi^L
\bigr] -
\bigl[{\psi^R}^\top X \,\psi^L + 
{\overline{\psi}^L}^\top X^\dagger\,{\overline{\psi}^R}
\bigr] .
\label{eqn:RenChar3}
\label{eqn:RenCharLagrangedichte1}
\end{eqnarray}
The mass matrix
\begin{eqnarray}
X & = & \left(\begin{array}{cc}
M_2 & \sqrt{2}\,M_W\,\sin\beta\\
\sqrt{2}\,M_W\,\cos\beta & \mu
\end{array}\right)
\label{eqn:RenChar1a}
\end{eqnarray}
can be diagonalized by two unitary matrices $U$ and $V$,
yielding the tree-level chargino mass eigenstates
\begin{eqnarray}
\chi^R_j = U_{jk}\,\psi^R_k , \quad
\chi^L_j = V_{jk}\,\psi^L_k , \quad
U^\ast\,X\,V^\dagger = 
\left(\begin{array}{cc}
m_{\chrg{1}} & 0 \\
0 & m_{\chrg{2}}
\end{array}\right).\;\;
\label{eqn:RenChar2}
\end{eqnarray}
The tree-level definition of the corresponding 
chargino Dirac spinors $\chrg{i}$ $(i=1,2)$
is then given by 
\begin{eqnarray}
\chrg{i} = \left(\begin{array}{c}
\chi^L_i \\ \overline{\chi}^R_i
\end{array}\right) .
\label{eqn:CharDirac}
\end{eqnarray}
The squares of the  tree-level masses $m_{\chrg{1}}$ and $m_{\chrg{2}}$ 
arise as  the eigenvalues of the hermitian matrix
$XX^\dagger$,
\begin{eqnarray}
m^2_{\chrg{1},\chrg{2}} &=& \frac{1}{2}\, \bigg\{ 
       M_2^2 + |\mu|^2 + 2M_W^2 \mp \Big[ (M_2^2-|\mu|^2)^2 \nonumber \\ 
&&{}+ 4M_W^4\cos^2 2\beta
       + 4M_W^2(M_2^2+|\mu|^2+2\,{\rm Re}(\mu)\, M_2\, \sin 2\beta) 
       \Big]^{\frac{1}{2}} \bigg\}~.
\label{eqn:Charmasse}
\end{eqnarray}

\subsection{Renormalization of the chargino sector}
\label{subsec:RenCharginos}

Starting from the chargino Lagrangian (\ref{eqn:RenCharLagrangedichte1}),
we introduce renormalization constants for the
mass matrix $X$ and for the fields $\psi^L$, $\psi^R$ by the
transformation
\begin{eqnarray}
X & \to & X + \delta X \nonumber \\
\psi^L & \to & \left(1+\lfrac{\delta Z^L}{2}\right)\,\psi^L \nonumber \\
\psi^R & \to & \left(1+\lfrac{\delta Z^R}{2}\right)\,\psi^R\;\;.
\label{eqn:RvRChar2}
\end{eqnarray}
The matrix $\delta X$ is made of the counterterms for the 
parameters in the 
mass matrix $X$ in~(\ref{eqn:RenChar1a}),
\begin{eqnarray}
\delta X & = & \left(\begin{array}{cc}
\delta M_2 & \sqrt{2}\,\delta\big(M_W\,\sin\beta\big)\\
\sqrt{2}\,\delta\big(M_W\,\cos\beta\big) & \delta\mu
\end{array}\right)\;\;.
\label{eqn:RenChar1aa}
\end{eqnarray}
The matrix-valued field-renormalization constants $\delta Z^L$ and
$\delta Z^R$ are chosen diagonal. 
This minimal set of renormalization constants 
is sufficient to render 
both $S$-matrix elements 
and Green functions for charginos  finite
\cite{HKS2000}.

For later convenience, we  define in a next step the
one-loop versions of the transformations~(\ref{eqn:RenChar2})
of the fields
$\psi^{L,R}$ by 
\begin{eqnarray}
\chi^L & = & R_L\,\psi^L \nonumber \\
\chi^R & = & R_R\,\psi^R
\label{eqn:RvRChar4}
\end{eqnarray}
with general complex, non-singular 
$2\times 2$-matrices   
$R_L$ and $R_R$, which are UV finite.
In the one-loop expansion, 
these two matrices can be written as follows,
\begin{eqnarray}
R_L & = & \left( 1+\lfrac{\delta Z^V}{2}\right)\,V \, , 
\label{eqn:RvRChar5a} \nonumber \\
R_R & = & \left( 1+\lfrac{\delta Z^U}{2}\right)\,U \, ,
\label{eqn:RvRChar5}
\end{eqnarray}
where $U$ and $V$ are the unitary matrices from (\ref{eqn:RenChar2})
and $\delta Z^U$, $\delta Z^V$ are general complex
$2\times 2$-matrices of one-loop order.
By combining the field transformations 
(\ref{eqn:RvRChar2}) and
(\ref{eqn:RvRChar4}), (\ref{eqn:RvRChar5}) one finds
\begin{eqnarray}
\psi^L & \to & \left(1+\lfrac{\delta Z^L}{2}\right)V^\dagger
\left(1-\lfrac{\delta Z^V}{2}\right)\,\chi^L 
= \left(V^\dagger + \lfrac{\delta
  Z^L}{2}V^\dagger-V^\dagger\lfrac{\delta Z^V}{2} \right)\,\chi^L \, ,
\nonumber \\ 
\psi^R & \to & \left(1+\lfrac{\delta Z^R}{2}\right)U^\dagger
\left(1-\lfrac{\delta Z^U}{2}\right)\,\chi^R
= \left(U^\dagger + \lfrac{\delta
  Z^R}{2}U^\dagger-U^\dagger\lfrac{\delta Z^U}{2} \right)\,\chi^R \, ,
\label{eqn:RvRChar6}
\end{eqnarray}
which shows that in the renormalized 
MSSM Lagrangian $\delta Z^L$ and $\delta Z^V$ can
only occur in the combination $\lfrac{\delta
  Z^L}{2}V^\dagger-V^\dagger\lfrac{\delta Z^V}{2}$, whilst 
$\delta Z^R$ and $\delta Z^U$ always combine to
$\lfrac{\delta Z^R}{2}U^\dagger-U^\dagger\lfrac{\delta
Z^U}{2}$. 
Hence, actually only $4$ complex renormalization constants
are available for each $L$ and $R$ part.
In order to eliminate the redundant parameters we define
new field-renormalization constants
\begin{eqnarray}
\delta\tilde{Z}^L & = & 
V\big[\delta Z^L V^\dagger-V^\dagger\delta Z^V\big] 
= V \delta Z^L V^\dagger-\delta Z^V \, ,
\nonumber \\
\delta\tilde{Z}^R & = & 
U\big[\delta Z^R U^\dagger-U^\dagger\delta Z^U\big]
= U \delta Z^R U^\dagger- \delta Z^U \, ,
\label{eqn:RvRCharA6}
\end{eqnarray}
which are now general complex $2\times 2$-matrices.

Applying the transformations (\ref{eqn:RvRChar6}) for the fields and 
(\ref{eqn:RenChar1a}) for the parameters
to the
Lagrangian (\ref{eqn:RenCharLagrangedichte1}) yields the Born and
the counterterm parts. 
After a Fourier transformation they read, with 4-component spinors
and the projectors
$\omega_{L,R}=(1 \mp \gamma_5)/2$ :
\begin{subequations}
\begin{eqnarray}
\mathcal{L}_{\rm Born}  & = &
\overline{\chrg{i}}\,\big[\ps\,\delta_{ij} 
- \omega_L\,(U^\ast X V^\dagger)_{ij} 
- \omega_R\,(V X^\dagger U^\top)_{ij}
\big]\,\chrg{i} \, ,
\\[0.4cm]
\mathcal{L}_{\rm CT} & = &
\overline{\chrg{i}}\,\ps
\Bigl[
\lfrac{1}{2}\left({\delta\tilde{Z}^R}^\ast+ {\delta\tilde{Z}^R}^\top
\right)_{ij}\,\omega_R
+ 
\lfrac{1}{2}\left(\delta\tilde{Z}^L+ {\delta\tilde{Z}^L}^\dagger
\right)_{ij}\,\omega_L\Bigr]\,\chrg{j}  \nonumber\\
&&{}-
\overline{\chrg{i}}
\Biggl[
\left[
U^\ast\delta X V^\dagger+
\lfrac{1}{2}\,{\delta\tilde{Z}^R}^\top U^\ast X V^\dagger +
\lfrac{1}{2}\,U^\ast X V^\dagger {\delta\tilde{Z}^L}
\right]_{ij}\,\omega_L
\nonumber \\
&&\hspace{1cm}  
+
\left[
V\delta X^\dagger U^\top+
\lfrac{1}{2}\,V X^\dagger U^\top {\delta\tilde{Z}^R}^\ast+
\lfrac{1}{2}\,{\delta\tilde{Z}^L}^\dagger V X^\dagger U^\top
\right]_{ij}\,\omega_R 
\Biggr] \chrg{j} \, .
\label{eqn:RvRChar7b}
\end{eqnarray}
\end{subequations}

The renormalized self-energies
$\hat{\Sigma}_{ij}(p)$
for the chargino system
are given by the unrenormalized self-energies 
$\Sigma_{ij}(p)$ plus the corresponding counterterms, 
obtained as derivatives of 
the  counterterm Lagrangian (\ref{eqn:RvRChar7b})
with respect to the fields $\overline{\chrg{i}}$ and $\chrg{j}$,
\begin{eqnarray}
\hat{\Sigma}_{ij}(p) & = & 
\Sigma_{ij}(p) + 
\frac{\partial}{\partial \chrg{j}}\frac{\partial}{\partial
  \overline{\chrg{i}}}\; \mathcal{L}_{\rm CT}\;\;.
\label{eqn:RvRChar9a}
\end{eqnarray}
Thus, by using the decomposition into Lorentz covariants
\begin{eqnarray}
\Sigma_{ij}(p) & = &
\ps\,\omega_L\Sigma_{ij}^L(p^2) 
+\ps\,\omega_R\Sigma_{ij}^R(p^2)
+ \omega_L\Sigma_{ij}^{SL}(p^2)
+ \omega_R\Sigma_{ij}^{SR}(p^2) \, ,
\label{eqn:CharginoSelbstenergie1}
\end{eqnarray}
one immediately obtains 
\begin{subequations}
\label{eqn:RvRChar10}
\begin{eqnarray}
\hat{\Sigma}_{ij}^R(p^2) & = & \Sigma_{ij}^R(p^2)+
\lfrac{1}{2}\left({\delta\tilde{Z}^R}^\ast+ {\delta\tilde{Z}^R}^\top
\right)_{ij}
\\
\hat{\Sigma}_{ij}^L(p^2) & = & \Sigma_{ij}^L(p^2)+
\lfrac{1}{2}\left(\delta\tilde{Z}^L+ {\delta\tilde{Z}^L}^\dagger
\right)_{ij}
\\
\hat{\Sigma}_{ij}^{SR}(p^2) & = &\Sigma_{ij}^{SR}(p^2) -
\left[
V\delta X^\dagger U^\top+
\lfrac{1}{2}\,V X^\dagger U^\top {\delta\tilde{Z}^R}^\ast+
\lfrac{1}{2}\,{\delta\tilde{Z}^L}^\dagger V X^\dagger U^\top
\right]_{ij}
\\
\hat{\Sigma}_{ij}^{SL}(p^2)&=&\Sigma_{ij}^{SL}(p^2) -
\left[
U^\ast\delta X V^\dagger+
\lfrac{1}{2}\,{\delta\tilde{Z}^R}^\top U^\ast X V^\dagger +
\lfrac{1}{2}\,U^\ast X V^\dagger {\delta\tilde{Z}^L}
\right]_{ij} 
\end{eqnarray}
\end{subequations}
for the scalar coefficients.

\section{Neutralinos}
\label{sec:Neutralinos}
\subsection{Lagrangian and mass eigenstates}

The bilinear, non-interacting, part of the neutralino Lagrangian
derived from (\ref{eqn:MSSMLkin}) and 
(\ref{eqn:MSSMPhotonWZ3a}) can be written in the conventional 
compact form
\begin{eqnarray}
\mathcal{L}_{\rm n} & = 
\lfrac{i}{2}\bigl[
{\psi^0}^\top\sigma^\mu\partial_\mu\,\overline{\psi}^0 +
{\overline{\psi}^0}^\top\overline{\sigma}^\mu\partial_\mu\,\psi^0
\bigr] 
-\lfrac{1}{2}
\bigl[{\psi^0}^\top Y \,\psi^0 + 
{\overline{\psi}^0}^\top Y^\dagger\,{\overline{\psi}^0}
\bigr] \, ,
\label{eqn:RenNeutr3}
\label{eqn:RenNeutrLagrangedichte1}
\end{eqnarray}
where the Weyl spinors of the neutral field components
are arranged as quadruples
\begin{eqnarray}
{\psi^0}^\top = \bigl(
-i\lambda',\,-i\lambda^3,\,\psi_{H_1}^1,\,\psi_{H_2}^2
\bigr) \, , \quad
{\overline{\psi}^0}^\top = \bigl(
i\overline{\lambda}',\,i\overline{\lambda}^3,\,
\overline{\psi}_{H_1}^1,\,\overline{\psi}_{H_2}^2
\bigr) \, .
\label{eqn:RenNeutr1}
\end{eqnarray}
The symmetric mass matrix, with $s_W=\sin\theta_W, \, 
c_W=\cos\theta_W$ for the electroweak mixing angle,
\begin{eqnarray}
Y & = & \left(\begin{array}{cccc}
M_1 & 0 & -M_Z\,s_W\,\cos\beta & M_Z\,s_W\,\sin\beta \\
0 & M_2 &  M_Z\,c_W\,\cos\beta & -M_Z\,c_W\,\sin\beta \\
-M_Z\,s_W\,\cos\beta & M_Z\,c_W\,\cos\beta & 0 & -\mu \\
M_Z\,s_W\,\sin\beta & -M_Z\,c_W\,\sin\beta & -\mu & 0
\end{array}\right) \, ,
\label{eqn:RenNeutr1a}
\end{eqnarray}
can be diagonalized with the help of a unitary $4 \times 4$ matrix $N$,
\begin{eqnarray}
N^\ast Y N^\dagger & = & M^D = \left(\begin{array}{cccc}
m_{1} & 0 & 0 & 0 \\ 
0 & m_{2} & 0 & 0 \\ 
0 & 0 & m_{3} & 0 \\ 
0 & 0 & 0 & m_{4} \\ 
\end{array}\right) \, ,
\label{eqn:RenNeutr2a}
\end{eqnarray}
yielding the neutralino mass eigenstates as linear combinations 
of the fields in (\ref{eqn:RenNeutr1}): 
\begin{eqnarray}
\chi^0_i = N_{ij}\,\psi^0_j \, , \quad \quad
\ntrl{i} = \left(\begin{array}{c}
\chi^0_i \\ \overline{\chi}^0_i
\end{array}\right) . 
\quad \quad
\label{eqn:RenNeutr2}
\end{eqnarray}
In the 4-component notation, 
the neutralino Majorana spinors
are denoted by $\ntrl{i}$ (with $i=1,\ldots,4)$.

\subsection{Renormalization of the neutralino sector}

Starting from the neutralino Lagrangian 
(\ref{eqn:RenNeutrLagrangedichte1}),
the mass-matrix $Y$ and the  fields $\psi^0$ 
are -- in analogy to the chargino case --
transformed by the following parameter and field renormalization,
\begin{eqnarray}
Y & \to & Y + \delta Y \nonumber\\
\psi^0 & \to & \left(1+\lfrac{\delta Z^0}{2}\right)\,\psi^0\;\;,
\label{eqn:RvRNeutr2b}\label{eqn:RvRNeutr2}
\end{eqnarray}
The counterterm matrix $\delta Y$ contains, besides those 
parameter counterterms already present in (\ref{eqn:RenChar1aa}),
the counterterms for the $Z$ mass and for the electroweak mixing angle,
respectively.
The matrix-valued renormalization constant $\delta Z^0$ is chosen diagonal.
As in the chargino case, this is sufficient for UV finiteness.
In the next step, for later convenience, 
we define the one-loop version of
 $\chi^0$ as a linear transformation  of the renormalized
fields $\psi^0$ via
\begin{eqnarray}
\chi^0 & = & R\,\psi^0
\label{eqn:RvRNeutr4}
\end{eqnarray}
with a complex, non-singular and UV-finite
$4\times 4$-matrix $R$.
For the one-loop expansion we can write
\begin{eqnarray}
R & = & \left( 1+\lfrac{\delta Z^N}{2}\right)\,N  \, ,
\label{eqn:RvRNeutr5}
\end{eqnarray}
where $N$ is the unitary matrix from (\ref{eqn:RenNeutr2a})
whereas $\delta Z^N$ is a general complex $4\times 4$-matrix of
one-loop order.

Performing the renormalization and redefinition of the neutralino fields
according to (\ref{eqn:RvRNeutr2b}), (\ref{eqn:RvRNeutr4}) and
(\ref{eqn:RvRNeutr5}), one ends up with the following net substitution
\begin{eqnarray}
\psi^0 & \to & \left(1+\lfrac{\delta Z^0}{2}\right)N^\dagger
\left(1-\lfrac{\delta Z^N}{2}\right)\,\chi^0 
= \left(N^\dagger + \lfrac{\delta
  Z^0}{2}N^\dagger-N^\dagger\lfrac{\delta Z^N}{2} \right)\,\chi^0\;\;.
\label{eqn:RvRNeutr6}
\end{eqnarray}
This makes obvious that the renormalization constants 
$\delta Z^0$ and $\delta Z^N$ can only occur in the
combination $\lfrac{\delta Z^0}{2}N^\dagger-N^\dagger\lfrac{\delta
Z^N}{2}$ throughout the MSSM Lagrangian.
In order to avoid redundances we define new field-renormalization constants
\begin{eqnarray}
\delta\tilde{Z}^0 & = & N\big[
\delta Z^0 N^\dagger-N^\dagger\delta Z^N\big] =
 N \delta Z^0 N^\dagger-\delta Z^N \, ,
\label{eqn:RvRNeutr7}
\end{eqnarray}
in analogy to those of the chargino case.
$\delta\tilde{Z}^0$ is now a general complex
$4\times 4$-matrix. 

Expressing the Lagrangian (\ref{eqn:RenNeutr3}) in terms of the new
fields $\chi^0$ and substituting the mass matrix according to 
(\ref{eqn:RvRNeutr2b}) yield
the Born and the counterterm Lagrangian for the neutralinos. 
Using the 4-component 
Majorana spinors $\ntrl{j}$
from (\ref{eqn:RenNeutr2}) they read,
after a Fourier transformation:
\begin{subequations}
\begin{eqnarray}
\mathcal{L}_{\rm Born}  & = &
\lfrac{1}{2}\;\overline{\ntrl{i}}\,\bigl[
\ps\,\delta_{ij} - 
\big(N^\ast\,Y\,N^\dagger\big)_{ij}\,\omega_L  - 
\big(N\,Y^\dagger\,N^\top\big)_{ij}\,\omega_R
\bigr]\,\ntrl{j} \, , \\[0.4cm]
\mathcal{L}_{\rm CT} & = &
\lfrac{1}{2}\,
\overline{\ntrl{i}}\,\ps\,
\Bigl[
\lfrac{1}{2}\left({\delta\tilde{Z}^0}^\ast + 
{\delta\tilde{Z}^0}^\top \right)_{ij} \omega_R  
+ \lfrac{1}{2}\left(\delta\tilde{Z}^0 + 
{\delta\tilde{Z}^0}^\dagger \right)_{ij} \omega_L
\Bigr]\,\ntrl{j}  \nonumber\\
&&{}-
\lfrac{1}{2}\,\overline{\ntrl{i}}\,\Bigl[
\left( N^\ast\delta Y N^\dagger + 
\lfrac{ {\delta\tilde{Z}^0}^\top N^\ast Y N^\dagger +  
N^\ast Y N^\dagger {\delta\tilde{Z}^0}
}{2} \right)_{ij}
\omega_L
\nonumber \\ && \hspace{1.4cm} +\,  
\left( N {\delta Y}^\dagger N^\top+
\lfrac{ N Y^\dagger N^\top {\delta\tilde{Z}^0}^\ast +
{\delta\tilde{Z}^0}^\dagger N Y^\dagger N^\top  
}{2} \right)_{ij}
\omega_R
\Bigr]\,\ntrl{j} \, .
\label{eqn:RvRNeutr9}
\end{eqnarray}
\end{subequations}

The neutralino self-energies are decomposed into Lorentz covariants 
as given in (\ref{eqn:CharginoSelbstenergie1}). The renormalized 
self-energies are obtained by adding the appropriate counterterms
following from (\ref{eqn:RvRNeutr9}) with the help of 
(\ref{eqn:RvRChar9a}), yielding 
\begin{subequations}
\label{eqn:RvRNeutr11}
\begin{eqnarray}
\hat{\Sigma}_{ij}^R(p^2) & = & \Sigma_{ij}^R(p^2)+
\lfrac{1}{2}\left({\delta\tilde{Z}^0}^\ast + 
{\delta\tilde{Z}^0}^\top \right)_{ij}
\\
\hat{\Sigma}_{ij}^L(p^2) & = & \Sigma_{ij}^L(p^2)+
\lfrac{1}{2}\left(\delta\tilde{Z}^0 + 
{\delta\tilde{Z}^0}^\dagger \right)_{ij}
\\
\hat{\Sigma}_{ij}^{SR}(p^2) & = &\Sigma_{ij}^{SR}(p^2) -
\left( N {\delta Y}^\dagger N^\top+
\lfrac{ N Y^\dagger N^\top {\delta\tilde{Z}^0}^\ast +
{\delta\tilde{Z}^0}^\dagger N Y^\dagger N^\top  
}{2} \right)_{ij}
\\
\hat{\Sigma}_{ij}^{SL}(p^2)&=&\Sigma_{ij}^{SL}(p^2) -
\left( N^\ast\delta Y N^\dagger + 
\lfrac{ {\delta\tilde{Z}^0}^\top N^\ast Y N^\dagger +  
N^\ast Y N^\dagger {\delta\tilde{Z}^0}
}{2} \right)_{ij}\;\;\;.
\end{eqnarray}
\end{subequations}

Since neutralinos are Majorana fermions,  the appropriate renormalized
self-energies have to obey the relations
\begin{equation}
\hat{\Sigma}^L_{ij}(p^2) = \hat{\Sigma}^R_{ji}(p^2) \, ,\quad 
\hat{\Sigma}^{SR}_{ij}(p^2) = \hat{\Sigma}^{SR}_{ji}(p^2) \, ,\quad
\hat{\Sigma}^{SL}_{ij}(p^2)  =  \hat{\Sigma}^{SL}_{ji}(p^2) \, ,
\label{eqn:RvRNeutrBed05}
\end{equation}
which are in accordance with the
structure of the counterterms in 
(\ref{eqn:RvRNeutr11}).

\section{On-shell conditions}

The propagators of the charginos and neutralinos explicitly depend on
the mass parameters $\mu$, $M_2$ and $M_1$ of the MSSM Lagrangian. We
now define the on-shell values of these parameters through the pole
positions of the propagators, which correspond to the physical masses of
the charginos and neutralinos. 

In addition we require for both charginos and neutralinos that the
matrix of the renormalized one-particle-irreducible two-point vertex
functions $\hat{\Gamma}^{(2)}_{ij}$ becomes diagonal for on-shell
external momenta. 
This fixes the non-diagonal entries of the field-renormalization
matrices; their diagonal entries are determined by normalizing the
residues of the propagators. 

The formulae of the previous sections are general enough to accommodate
also complex MSSM parameters giving rise to intrinsic CP violation. 
In the following discussion we restrict ourselves to the simpler case of 
the CP-conserving MSSM with real parameters.
 
\subsection{Charginos}

In the case of CP conservation the on-shell renormalization conditions
for the chargino sector correspond to the following relations between
the renormalized self-energies (\ref{eqn:RvRChar10}), with $i,j=1,2$
[the operation $\ReTilde$ replaces the momentum integral in the following
term by its real part but does not change other complex coefficients]:
\begin{subequations}
\label{eqn:RvRCharBed10}
\begin{eqnarray}
U^\ast X V^\dagger & = & {\rm diag}(m_{\chrg{1}},m_{\chrg{2}})
\label{eqn:RvRCharBed10a} \\ 
\nonumber  \\
m_{\chrg{j}} \ReTilde\hat{\Sigma}_{ij}^R(m_{\chrg{j}}^2) +
\ReTilde\hat{\Sigma}_{ij}^{SL}(m_{\chrg{j}}^2)
& = & 0 \nonumber \\
m_{\chrg{j}} \ReTilde\hat{\Sigma}_{ij}^L(m_{\chrg{j}}^2) +
\ReTilde\hat{\Sigma}_{ij}^{SR}(m_{\chrg{j}}^2)
& = & 0 \label{eqn:RvRCharBed10b}\\ 
\nonumber \\
\ReTilde\hat{\Sigma}^L_{ii}(m_{\chrg{i}}^2) 
+ 2m_{\chrg{i}} 
\ReTilde\hat{\Sigma}^{SL}_{ii}\,\!'(m_{\chrg{i}}^2)
\hspace{3.0cm} \nonumber \\
{}+
m_{\chrg{i}}^2 \left(
 \ReTilde\hat{\Sigma}^L_{ii}\,\!'(m_{\chrg{i}}^2)
+ \ReTilde\hat{\Sigma}^R_{ii}\,\!'(m_{\chrg{i}}^2)
\right)
& = & 0\;\;.
\end{eqnarray}
\end{subequations}
The diagonal $(i=j)$ equations in (\ref{eqn:RvRCharBed10b}) ensure that
the positions of the propagator poles are not shifted by the renormalized
self-energies. This means that the relations between the chargino pole
masses $m_{\chrg{i}}$ and the MSSM parameters have the same form as in
lowest order, also at the one-loop level. 

Inserting the renormalized chargino self-energies (\ref{eqn:RvRChar10})
into the ten equations above and solving for the renormalization
constants one obtains the explicit expressions 
\begin{subequations}
\label{eqn:RvRCharRenKonst2}
\begin{eqnarray}
\delta M_2 & = &
\Bigl[U_{22} V_{22} \Big(
m_{\chrg{1}}\big[\ReTilde\Sigma_{11}^L(m_{\chrg{1}}^2)+ 
\ReTilde\Sigma_{11}^R(m_{\chrg{1}}^2)\big] 
+2 \ReTilde\Sigma_{11}^{SL}(m_{\chrg{1}}^2)
\Big) \nonumber \\
&& \hspace{0 cm}
{}- U_{12} V_{12}\Big(
m_{\chrg{2}}\big[\ReTilde\Sigma_{22}^L(m_{\chrg{2}}^2)+ 
\ReTilde\Sigma_{22}^R(m_{\chrg{2}}^2)\big] 
+2 \ReTilde\Sigma_{22}^{SL}(m_{\chrg{2}}^2)
\Big) \nonumber \\
&& \hspace{0 cm}
{} + 2\,\big(U_{12}U_{21}-U_{11}U_{22}\big)V_{12}V_{22}\,\delta(\sqrt{2}M_W\sin\beta)
\nonumber \\ &&
{} + 2\,U_{12}U_{22}\big(V_{12}V_{21}-V_{11}V_{22}\big)\,\delta(\sqrt{2}M_W\cos\beta)\Bigr] /\Delta \, ,
\label{eqn:RvRCharRenKonst2a}
\end{eqnarray}
\begin{eqnarray}
\delta \mu & = & 
\Bigl[U_{11}V_{11}\Big(
m_{\chrg{2}}\big[\ReTilde\Sigma_{22}^L(m_{\chrg{2}}^2)+
\ReTilde\Sigma_{22}^R(m_{\chrg{2}}^2)\big]
+2 \ReTilde\Sigma_{22}^{SL}(m_{\chrg{2}}^2)
\Big)\nonumber \\
&& \hspace{0cm}
{}- U_{21} V_{21} \Big(
m_{\chrg{1}}\big[\ReTilde\Sigma_{11}^L(m_{\chrg{1}}^2)+
\ReTilde\Sigma_{11}^R(m_{\chrg{1}}^2)\big]
+2 \ReTilde\Sigma_{11}^{SL}(m_{\chrg{1}}^2)
\Big) \nonumber \\
&& \hspace{0cm}
{} + 2\,U_{11}U_{21}\big(V_{12} V_{21} - V_{11} V_{22}
\big)\,\delta(\sqrt{2}M_W\sin\beta)
\nonumber \\ &&
{} + 2\,\big( U_{12} U_{21} - U_{11}U_{22}\big) V_{11}
V_{21}\,\delta(\sqrt{2}M_W\cos\beta)\Bigr]  /\Delta \, ,
\label{eqn:RvRCharRenKonst2b}
\end{eqnarray}
with 
$\quad \Delta =  2(U_{11}U_{22}V_{11}V_{22}- U_{12}U_{21}V_{12}V_{21})$; 
\\[0.3cm]
\begin{eqnarray}
\delta \tilde{Z}^L_{ii} & = & 
-\ReTilde\Sigma_{ii}^L(m_{\chrg{i}}^2)-
m_{\chrg{i}}^2\big[
\ReTilde\Sigma_{ii}^{L'}(m_{\chrg{i}}^2)+
\ReTilde\Sigma_{ii}^{R'}(m_{\chrg{i}}^2)
\big] - 
 2\,m_{\chrg{i}}\,\ReTilde\Sigma_{ii}^{SL'}(m_{\chrg{i}}^2) \, ,
\nonumber \\[0.4cm]
\delta \tilde{Z}^L_{ij} & = & 
\lfrac{2}{m_{\chrg{i}}^2-m_{\chrg{j}}^2} \, \cdot \,\Big[
m_{\chrg{j}}^2\,\ReTilde\Sigma_{ij}^{L}(m_{\chrg{j}}^2) +
m_{\chrg{i}}\,m_{\chrg{j}}\,\ReTilde\Sigma_{ij}^{R}(m_{\chrg{j}}^2) +
m_{\chrg{i}}\,\ReTilde\Sigma_{ij}^{SL}(m_{\chrg{j}}^2)
\nonumber \\
& & \hspace{2.5cm}
{}+m_{\chrg{j}}\,\ReTilde\Sigma_{ji}^{SL}(m_{\chrg{j}}^2)
- m_{\chrg{i}}\,(U \delta X V^\top)_{ij} 
- m_{\chrg{j}}\,(U \delta X V^\top)_{ji}\Big] \, ,
\nonumber \\
\end{eqnarray}
\begin{eqnarray}
\delta \tilde{Z}^R_{ii} & = & 
-\ReTilde\Sigma_{ii}^R(m_{\chrg{i}}^2)-
m_{\chrg{i}}^2\big[
\ReTilde\Sigma_{ii}^{L'}(m_{\chrg{i}}^2)+
\ReTilde\Sigma_{ii}^{R'}(m_{\chrg{i}}^2)
\big] - 
2\,m_{\chrg{i}}\,\ReTilde\Sigma_{ii}^{SL'}(m_{\chrg{i}}^2) \, ,
\nonumber \\[0.4cm]
\delta \tilde{Z}^R_{ij} & = & 
\lfrac{2}{m_{\chrg{i}}^2-m_{\chrg{j}}^2}\, \cdot \,\Big[
m_{\chrg{j}}^2\,\ReTilde\Sigma_{ij}^{R}(m_{\chrg{j}}^2) +
m_{\chrg{i}}\,m_{\chrg{j}}\,\ReTilde\Sigma_{ij}^{L}(m_{\chrg{j}}^2) +
m_{\chrg{j}}\,\ReTilde\Sigma_{ij}^{SL}(m_{\chrg{j}}^2)
\nonumber \\
& & \hspace{2.5cm}
{}+m_{\chrg{i}}\,\ReTilde\Sigma_{ji}^{SL}(m_{\chrg{j}}^2)
- m_{\chrg{j}}\,(U \delta X V^\top)_{ij} 
- m_{\chrg{i}}\,(U \delta X V^\top)_{ji}\Big] \, .
\nonumber \\\label{eqn:RvRCharRenKonst2c}
\end{eqnarray}
\end{subequations}
So far, the renormalization of 
two of our MSSM parameters ($M_2$ and $\mu$) has been fixed 
by the chargino mass renormalization.

\subsection{Neutralinos}

The left-over mass-parameter of the MSSM Lagrangian
yet to be determined is the Bino mass $M_1$ of the neutralino sector.
In the on-shell strategy, it can be fixed, together with its counterterm,
by the on-shell mass renormalization 
of one of the four neutralino states, which we choose to be 
$m_{\ntrl{1}}$.

Moreover,  the additional matrix 
(\ref{eqn:RvRNeutr4}) in the field-renormalization constants 
allows one to impose the condition of having
diagonal renormalized 2-point vertex functions for each of
the neutralinos on-shell, i.e. $(i\neq j)$
\[ \hat{\Gamma}^{(2)}_{ij}(p) = 0
\quad  {\rm for\;  either} \quad 
 p^2 = m_{\ntrl{i}}^2 \quad  {\rm or} \quad
 p^2 = m_{\ntrl{j}}^2 \, . \]
This fixes the 12 non-diagonal entries of 
$\delta\tilde{Z}^0$. 
The remaining four diagonal entries
are determined by requiring unity for
the residues of the neutralino propagators.

In the case of CP-conservation this leads to the following conditions:
\begin{subequations}
\label{eqn:RvRNeutrBed2}
\begin{eqnarray}
\left(N Y N^\top\right)_{11} & = & m_{\ntrl{1}} \nonumber \\
N Y N^\top  =  {\rm diag}(m_1,m_2,m_3,m_4)&\equiv& M^D\hspace{0.9cm}\vphantom{i}
\label{eqn:RvRNeutrBed2a}\\ \nonumber \\
m_{\ntrl{j}} \ReTilde\hat{\Sigma}_{ij}^L(m_{\ntrl{j}}^2) +
\ReTilde\hat{\Sigma}_{ij}^{SL}(m_{\ntrl{j}}^2)
& = & 0 \label{eqn:RvRNeutrBed2b} \nonumber \\
\mbox{for  }(i\neq j) \vee (i=j=1)
\\ \nonumber \\
\ReTilde\hat{\Sigma}^L_{ii}(m_{\ntrl{i}}^2) +
2 m_{\ntrl{i}}^2\,
 \ReTilde\hat{\Sigma}^L_{ii}\,\!'(m_{\ntrl{i}}^2) 
+2 m_{\ntrl{i}}
 \ReTilde\hat{\Sigma}^{SL}_{ii}\,\!'(m_{\ntrl{i}}^2)
& = & 0 \;\;\;.
\label{eqn:RvRNeutrBed2c}
\end{eqnarray}
\end{subequations}
The value of $M_1$ is related to the  mass $m_{\ntrl{1}}$ of $\ntrl{1}$
by means of (\ref{eqn:RvRNeutrBed2a}). The condition
(\ref{eqn:RvRNeutrBed2b}) for $i=j=1$ ensures that this is also the pole
mass at the one-loop level. After this step, all eigenvalues
$m_j\;(j=1,\ldots,4)$ of the matrix $Y$ are known. 
However, only $m_1\equiv m_{\ntrl{1}}$ is equal to the pole mass. 
The other eigenvalues $m_{2,3,4}$ are the Born approximations of the
corresponding physical neutralino masses. They get corrections at the
one-loop level, as discussed in the next section.

Inserting the renormalized neutralino self-energies (\ref{eqn:RvRNeutr11})
and solving the equations~(\ref{eqn:RvRNeutrBed2}) for the
renormalization constants one finds the explicit expressions
\begin{subequations}
\label{eqn:RvRNtrlRenKonst1}
\begin{eqnarray}
\delta M_1 & = & \frac{1}{N_{11}^2}\Bigl[
2 N_{11}\big[N_{13}\,\delta(M_Z\sin\theta_W\cos\beta)-
N_{14}\,\delta(M_Z\sin\theta_W\sin\beta) \big] - \nonumber \\
&&\hspace{1cm}
N_{12}\big[
2 N_{13}\, \delta(M_Z\cos\theta_W\cos\beta)
-2 N_{14}\, \delta(M_Z\cos\theta_W\sin\beta)
+N_{12}\,\delta M_2\big] \nonumber \\
&& \hspace{1cm} + 2 N_{13} N_{14}\, \delta\mu + m_{\ntrl{1}} 
\ReTilde\Sigma^L_{11}(m_{\ntrl{1}}^2) + 
\ReTilde\Sigma^{SL}_{11}(m_{\ntrl{1}}^2) 
\Bigr] \, ,
\end{eqnarray}
\begin{eqnarray}
\delta \tilde{Z}^0_{ii} & = & 
-\ReTilde\Sigma_{ii}^L(m_{\ntrl{i}}^2)-
2 m_{\ntrl{i}}\big[
m_{\ntrl{i}}\, \ReTilde\Sigma_{ii}^{L'}(m_{\ntrl{i}}^2)+
\ReTilde\Sigma_{ii}^{SL'}(m_{\ntrl{i}}^2)
\big] \, ,  \\[0.4cm]
\delta \tilde{Z}^0_{ij} & = & 
\frac{2\big[
m_{\ntrl{j}}\, \ReTilde\Sigma_{ij}^{L}(m_{\ntrl{j}}^2)
+ \ReTilde\Sigma_{ij}^{SL}(m_{\ntrl{j}}^2)
- (N \delta Y N^\top)_{ij}
\big]
}{m_{\ntrl{i}} - m_{\ntrl{j}}}  \, .
\end{eqnarray}
\end{subequations}
Together with the renormalization constants from the gauge and Higgs
sector, outlined in the following subsection, the renormalization of the
2-point functions of the chargino/neut\-ral\-ino sector of the MSSM is
complete. Moreover, all  renormalization constants are now available to
determine all the counterterms required for one-loop calculations in the
neutralino--chargino sector of the MSSM. 

\subsection{Renormalization constants from other sectors}

The two sets of equations (\ref{eqn:RvRCharRenKonst2}) and
(\ref{eqn:RvRNtrlRenKonst1}) for the renormalization constants contain
explicitly the counterterms for the quantities $M_W$, $M_Z$, $\theta_W$
of the gauge sector and for $\beta$, which is a parameter of the Higgs
sector. 

In the on-shell scheme, the renormalization of the electroweak weak
mixing angle with $c_W = M_W/M_Z$ is deduced from the renormalization of
the $W$- and $Z$-boson masses, at the one-loop level via the relation 
\[ 
 \delta s_W^2 \, = \, c_W^2 \, \left(
    \frac{\delta M_Z^2}{M_Z^2} - 
    \frac{\delta M_W^2}{M_W^2} \right) \, .
\]
The on-shell counterterms for $M_W$ and $M_Z$ are given by
the transverse parts of the respective
vector-boson self-energies evaluated on their mass shell, 
\begin{eqnarray}
\delta M_W^2 & = & {\rm Re}\,  \Sigma_{WW}^{\rm trans}(M_W^2) \, ,
\label{eqn:OtherRCx1}
\nonumber \\[0.3cm]
\delta M_Z^2 & = & {\rm Re}\, \Sigma_{ZZ}^{\rm trans}(M_Z^2) \, .
\label{eqn:OtherRCx2}
\end{eqnarray}

Following \cite{Dabelstein95} we fix the renormalization constant
for $\tan\beta$ by the condition
\begin{eqnarray}
\frac{\delta\tan\beta}{\tan\beta} & = &
\frac{1}{2 M_Z \sin\beta\,\cos\beta}\cdot{\rm Im}\, 
\big[\ReTilde\Sigma_{A^0 Z}(M_A^2)\big]\;\;.
\label{eqn:OtherRCx3}
\end{eqnarray}
Another option, which has been applied in the recent version of 
{\it FeynHiggs}~\cite{FeynHiggs},
would be a $\overline{\rm MS}$ renormalization of $\tan\beta$.

\section{Neutralino masses}

The renormalization procedure presented in the last section assures
the neutralino fields not to mix with each other on their specific
mass-shell. Thus the one-loop corrected masses of the remaining three
neutralinos can simply be determined by finding those momenta $p_i^2 = 
m_{\ntrl{i}}^2$ which obey the relation
\begin{eqnarray}
\ReTilde\left[\hat\Gamma^{(2)}_{ii}(p_i)\right] \,u(p_i) = 0\;\;\;,
\label{eqn:RvRNeutrMasse1}
\end{eqnarray}
where 
\bea
\hat\Gamma^{(2)}_{ij}(p) & = & (\ps - m_i)\, \delta_{ij}
                     \, + \, \hat\Sigma_{ij}(p) \, , \nonumber \\[0.3cm]
\hat{\Sigma}_{ij}(p) & = &
\ps\,\omega_L\hat{\Sigma}_{ij}^L(p^2) 
+\ps\,\omega_R\hat{\Sigma}_{ij}^R(p^2)
+ \omega_L\hat{\Sigma}_{ij}^{SL}(p^2)
+ \omega_R\hat{\Sigma}_{ij}^{SR}(p^2) \, ,
\eea
are the renormalized neutralino two-point vertex-functions with the
self-energies (\ref{eqn:RvRNeutr11}) and $u(p_i)$ the $i$-neutralino
wave function in momentum space. At one-loop order, the condition
(\ref{eqn:RvRNeutrMasse1}) has the solution
\begin{eqnarray}
m_{\ntrl{i}} & = & 
m_i\left[1- \ReTilde\hat\Sigma_{ii}^L(m_i^2)\right]-
\ReTilde\hat\Sigma_{ii}^{SL}(m_i^2)
\label{eqn:RvRNeutrMasse7}
\end{eqnarray}
for the neutralino pole masses.
Inserting (\ref{eqn:RvRNeutr11}) for the renormalized self-energies
finally yields the neutralino masses in terms of the unrenormalized
neutralino self-energies and the renormalization constants, 
\begin{eqnarray}
m_{\ntrl{i}} & = & 
m_i\big[1-\ReTilde\Sigma_{ii}^L(m_i^2)\big]-\ReTilde\Sigma_{ii}^{SL}(m_i^2)
+\big(N \delta Y N^\top\big)_{ii}\;\;\;.
\label{eqn:RvRNeutrMasse8}
\end{eqnarray}
These masses are thus predictions arising from
$m_{\chrg{1}}, m_{\chrg{2}}, m_{\ntrl{1}}$ and 
depend in addition on the residual MSSM parameters that enter
the self-energies and the counterterms at the one-loop level.

\section{Numerical evaluation}

\subsection{Specification of $\mu$, $M_2$ and $M_1$}

In our on-shell approach, the pole masses of the two charginos,
$m_{\chrg{1}}, m_{\chrg{2}}$, and of one neutralino, $m_{\ntrl{1}}$, are
considered as input parameters, to specify the chargino/neutralino
Lagrangian in terms of physical quantities. 
This is equivalent to the specification of the parameters $\mu, M_1,
M_2$, which are related to the input masses in the same way as in lowest
order, as a consequence of the on-shell renormalization conditions. 

For given pole masses of the charginos, the values of $M_2$ and $\mu$
can be determined using equation (\ref{eqn:Charmasse}). Inverting those
relations one gets four solutions corresponding to different physical
scenarios, 
\begin{eqnarray}
  M_2 =a_{\pm}  \, , & \quad &
  \mu =\frac{a_{\pm}\cdot a_{\mp}^2}{2 M_W^2 \sin\beta\,\cos\beta +
  m_{\chrg{1}}\;m_{\chrg{2}}} \, ;
 \nonumber \\
  M_2 =b_{\pm} \, , & \quad &
  \mu =\frac{b_{\pm}\cdot b_{\mp}^2}{2 M_W^2 \sin\beta\,\cos\beta -
  m_{\chrg{1}}\;m_{\chrg{2}}} \, ;
  \label{eqn:CPInvOnShellRechnung2}
\end{eqnarray}
with the abbreviations
\begin{eqnarray}
  a_{\pm} & = & \lfrac{1}{\sqrt{2}}
  \sqrt{ m_{\chrg{1}}^2 + m_{\chrg{2}}^2 - 2 M_W^2 \,\pm\, c_{+}} \nonumber \\
  b_{\pm} & = & \lfrac{1}{\sqrt{2}}
  \sqrt{ m_{\chrg{1}}^2 + m_{\chrg{2}}^2 - 2 M_W^2 \,\pm\, c_{-}} \nonumber \\
  c_{\pm} & = & \sqrt{
    \big(m_{\chrg{1}}^2 + m_{\chrg{2}}^2 - 2 M_W^2\big)^2 - 4
     \big(m_{\chrg{1}}\;m_{\chrg{2}} \pm 2 M_W^2 \sin\beta\,\cos\beta\big)^2}\;\;.
  \label{eqn:CPInvOnShellRechnung3}
\end{eqnarray}

After selecting a specific $M_2$--$\mu$ configuration all entries in $Y$
from (\ref{eqn:RenNeutr1a}) are determined, except for $Y_{11}\equiv M_1$. 
The value of $M_1$ is obtained by the condition that the  eigenvalue $m_1$
of $Y$  coincides with the pole mass  $m_{\ntrl{1}}$. For the case of
real parameters the appropriate eigenvalue equation leads to the unique solution
\begin{eqnarray}
  \hspace{-1.4cm} M_1 & = & \Bigl[
  - M_2\mu M_Z^2 \sin\,2\beta + \big[\mu M_Z^2\sin
  2\beta - M_2\big(\mu^2+M_Z^2 s_W^2\big)\big]m_{\ntrl{1}} 
  \nonumber \\
&&{}\hspace{0.5cm}
 +\big[\mu^2+M_Z^2\big] m_{\ntrl{1}}^2 + M_2 m_{\ntrl{1}}^3 - m_{\ntrl{1}}^4
   \; \Bigr] \nonumber \\
&& {}\cdot \Bigl[
  \mu M_Z^2 c_W^2 \sin\,2\beta - M_2 \mu^2 + \big[\mu^2+M_Z^2
  c_W^2\big] m_{\ntrl{1}} + M_2 m_{\ntrl{1}}^2 -
  m_{\ntrl{1}}^3\Bigr]^{-1}\;\;.
  \label{eqn:CPInvOnShellRechnung6}
\end{eqnarray}
After this specification of the mass parameters, the Born-level
diagonalization matrices $U$, $V$ and $N$ can be calculated.

\subsection{Results and discussion}

The self-energies appearing in the counter terms $\delta\mu$, 
$\delta M_2$, $\delta M_1$, $\delta M_W^2$, $\delta M_Z^2$ and 
$\delta \tan\beta$ as well as the neutralino self-energies
in equation (\ref{eqn:RvRNeutrMasse8}) have been calculated with the
help of the program packages  {\tt FeynArts}, {\tt FormCalc} and 
{\tt LoopTools}  \cite{FeynArts}.  
For regularization we used the method of 'Constrained Differential
Renormalization' \cite{Pe99}. At the one-loop level this prescription
has been proven to be equivalent to 'Dimensional Reduction' \cite{Si79}, 
which is compatible with supersymmetry.

In the numerical evaluation, mixing between the fermion families has
been neglected. Moreover, we assume a common sfermion-mass scale
$m^2_{\{\tilde{q},\tilde{u}, \tilde{d}, \tilde{l},
  \tilde{e}\}}\equiv M_{\rm susy}^2$ 
for simplification. 
The gaugino parameters $M_1$ and $M_2$ are treated as independent.

All values for the masses in figures 1--3 have to be understood in units
of GeV. The input values taken for the gauge-boson masses are
$M_W  =  80.419\;{\rm GeV}, \, M_Z =  91.1882\;{\rm GeV}$. 
For the MSSM parameters, unless stated differently, the following values
have been used for the examples in the numerical presentation.
Thereby, the trilinear $A_{u,d,e}$ parameters are assumed universal
for the three generations.
\begin{equation*}
\begin{array}{rclcrclcrcl}
\hline \\
M_A & = & 150\;{\rm GeV}
&\;\hspace{1cm}\;&
M_{\rm susy} & = & 300\;{\rm GeV}
&\;\hspace{1cm}\;&
\tan\beta & = & 10 \\
A_{u} & = & 100\;{\rm GeV}
& &
A_{d} & = & 900\;{\rm GeV}
& &
A_{e} & = & 900\;{\rm GeV} \\ \\
\hline 
\end{array}
\end{equation*}

Before entering the presentation of our results we want to add a few
comments regarding other treatments of on-shell renormalization. The
scheme used in \cite{guasch} for the calculation of sfermion decays into
fermions is equivalent to the one specified here up to the treatment of
field renormalization. In our case, the effective field-renormalization
constants involve a finite one-loop redefinition of the diagonalization
matrices $U,V$ and $N$ [see (\ref{eqn:RvRCharA6}) and
(\ref{eqn:RvRNeutr7})] which allows a complete diagonalization of the
self-energy matrices on-shell, whereas in \cite{guasch} the $U,V,N$ are
not modified. Differences for the predicted neutralino masses in terms
of the input masses, however, are only of higher order and thus
accordingly very small. This was  confirmed also numerically by an
explicit comparison~\cite{guasch1}.\footnote{
We thank J. Guasch for the numerical comparison with
the results based on the scheme of \cite{guasch}.
}

In \cite{eberl} the field-renormalization constants are introduced also
in combination with  one-loop redefinitions of the matrices $U,V,N$. The
$Z$ factors, however, have been fixed by an independent prescription
adopted  from~\cite{kniehlpilaftsis}. 
In order to make this compatible with the on-shell renormalization
conditions, this requires a re-adjustment of the entries in the chargino
and neutralino mass matrices by finite shifts; otherwise $M_W$ and $M_Z$
in (\ref{eqn:RenChar1a}) and (\ref{eqn:RenNeutr1a}) would not be the
on-shell (pole) masses of $W$ and $Z$. 
As a consequence, the MSSM parameters in the mass matrices are in
general different at tree level and one-loop order. 
In particular, the relations between $M_2, \mu$ and the chargino pole
masses as well as between $M_1$ and the $\ntrl{1}$ pole mass are not the
tree-level relations but do contain additional terms of one-loop order.
For comparisons one therefore has to keep in mind that the values of the
formal parameters $\mu, M_1, M_2$ in~\cite{eberl} are different from
ours for the same physical situation, i.e.\ the physical masses
$m_{\chrg{1}},\, m_{\chrg{1}},\, m_{\ntrl{1}}$.
The masses of the three other neutralinos, however, when calculated
as observables from the same physical input, should be the same, up to
small terms of formally higher orders.    
The evaluation in \cite{eberl} was performed for the subclass of
fermion/sfermion-loop contributions only; hence, for a numerical
comparison, we had to turn off the non-(s)fermionic loop contributions
of our approach and found indeed good agreement in the calculated
neutralino masses for the examples given in \cite{eberl}.\footnote{
We thank H. Eberl for providing us with the detailed
numbers for the examples given in \cite{eberl}.
}

\subsubsection{Dependence of the neutralino masses on
                  $m_{\chrg{1}}$ and $m_{\chrg{2}}$}

\begin{figure}[H]
\begin{center}
\vspace{0.5cm}
\psfrag{MC1}{\footnotesize $m_{\chrg{2}}$}
\psfrag{MC2}{\footnotesize $m_{\chrg{1}}$}
\psfrag{M1}{\footnotesize $M_1$}
\psfrag{MC2170}{\footnotesize $m_{\chrg{1}}=170$ GeV}
\psfrag{MC2135}{\footnotesize $m_{\chrg{1}}=135$ GeV}
\psfrag{MC2100}{\footnotesize $m_{\chrg{1}}=100$ GeV}
\psfrag{MNeu1}{\footnotesize $m_{\ntrl{2}}$}
\psfrag{MNeu2}{\footnotesize $m_{\ntrl{2}}$}
\psfrag{MNeu3}{\footnotesize $m_{\ntrl{3}}$}
\psfrag{MNeu4}{\footnotesize $m_{\ntrl{4}}$}
\hfil
\epsfig{file=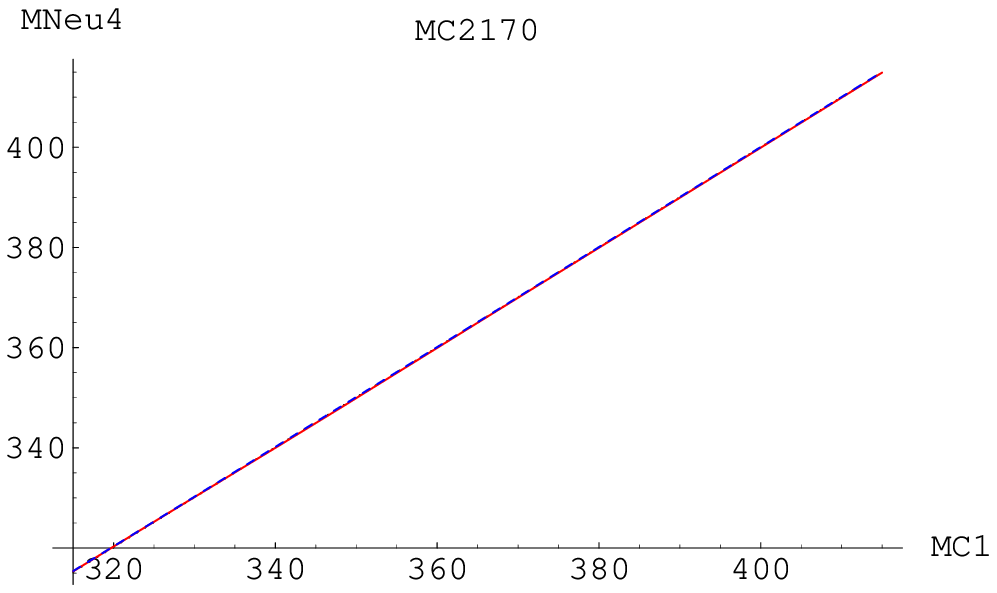,width=0.48\linewidth}
\hfil\\[0.3cm]
\epsfig{file=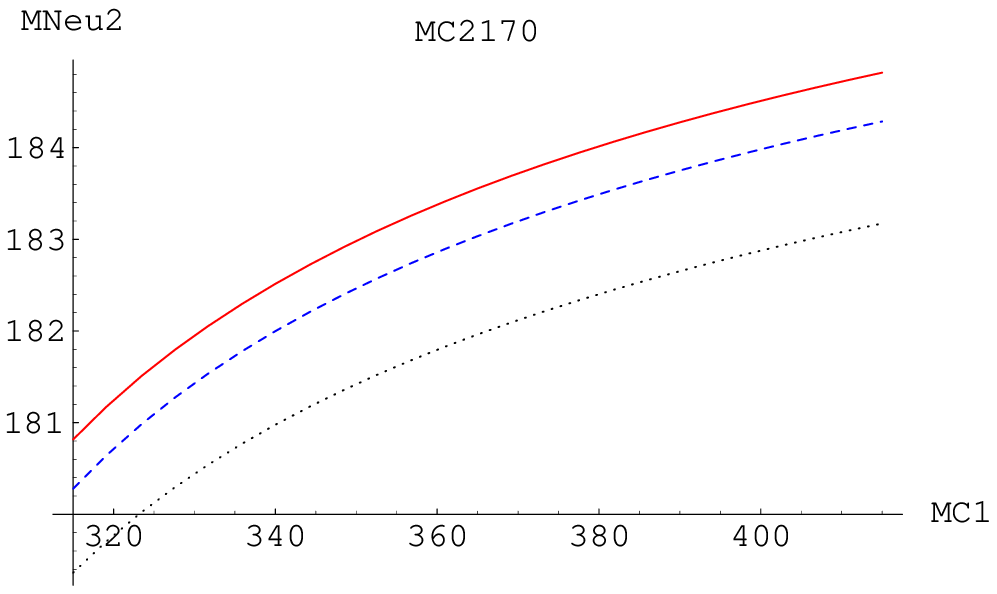,width=0.48\linewidth}
\hfil
\epsfig{file=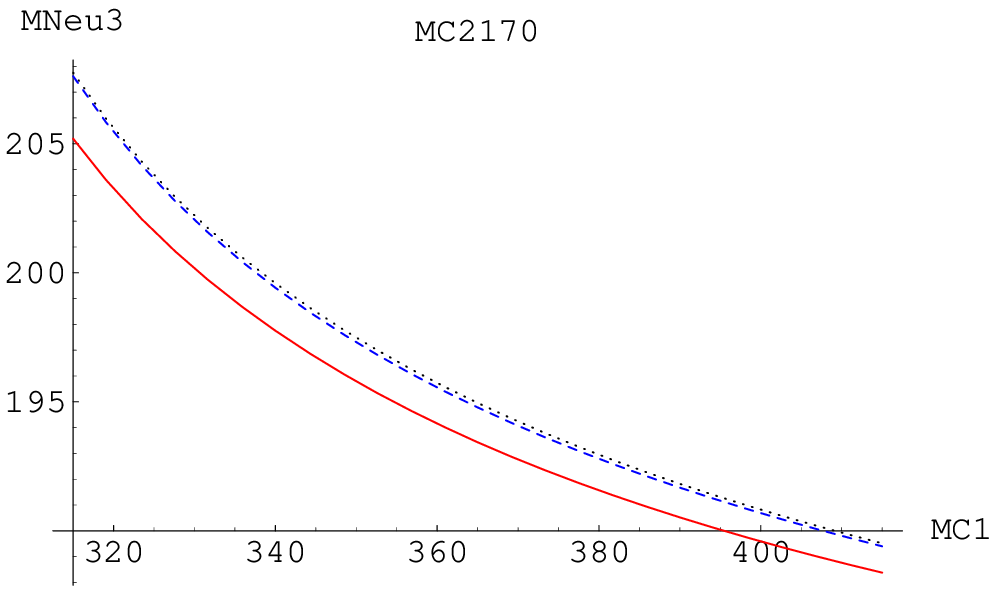,width=0.48\linewidth}
\\[0.7cm]
\epsfig{file=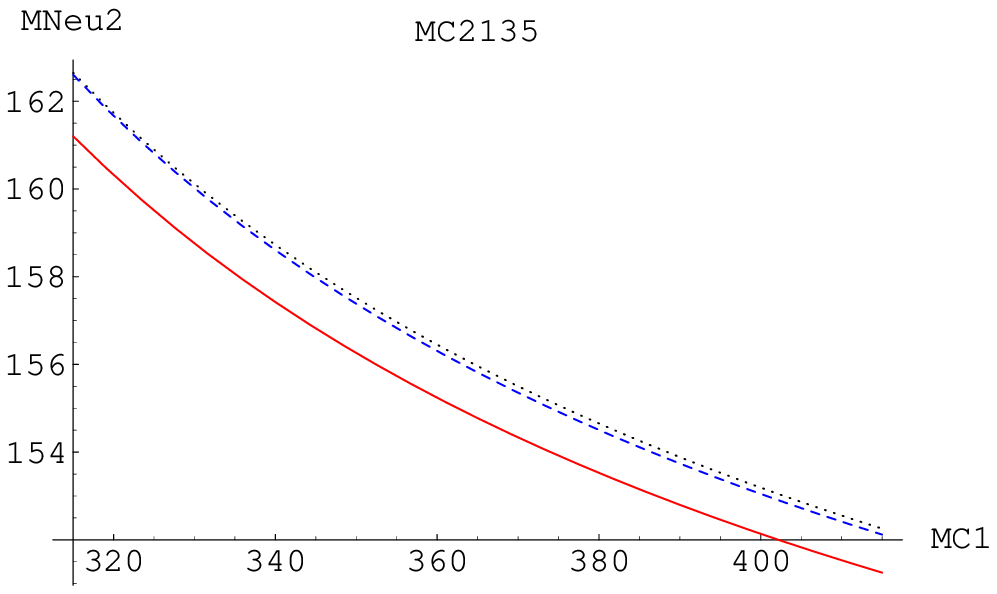,width=0.48\linewidth}
\hfil
\epsfig{file=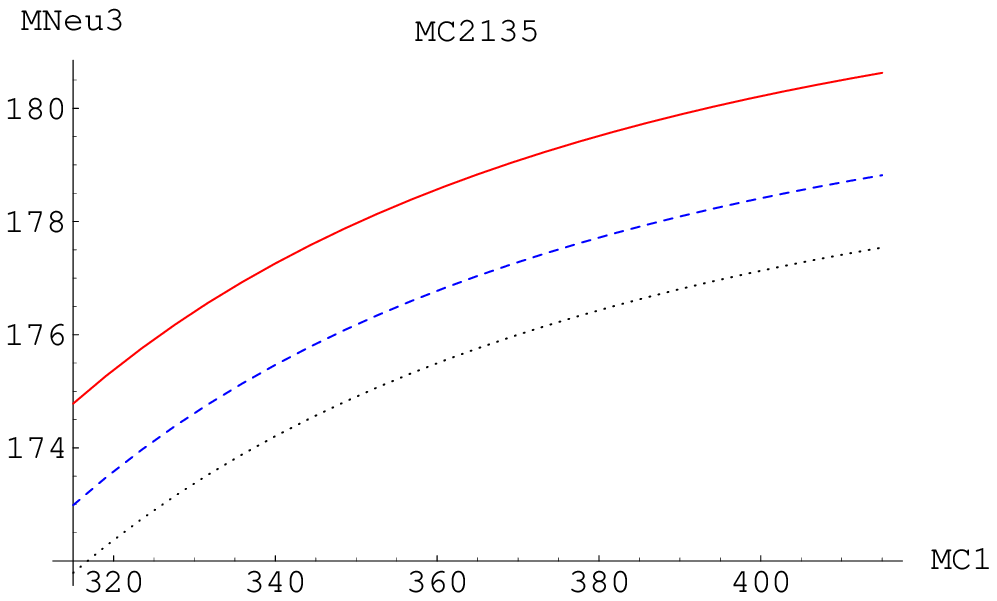,width=0.48\linewidth}
\\[0.7cm]
\epsfig{file=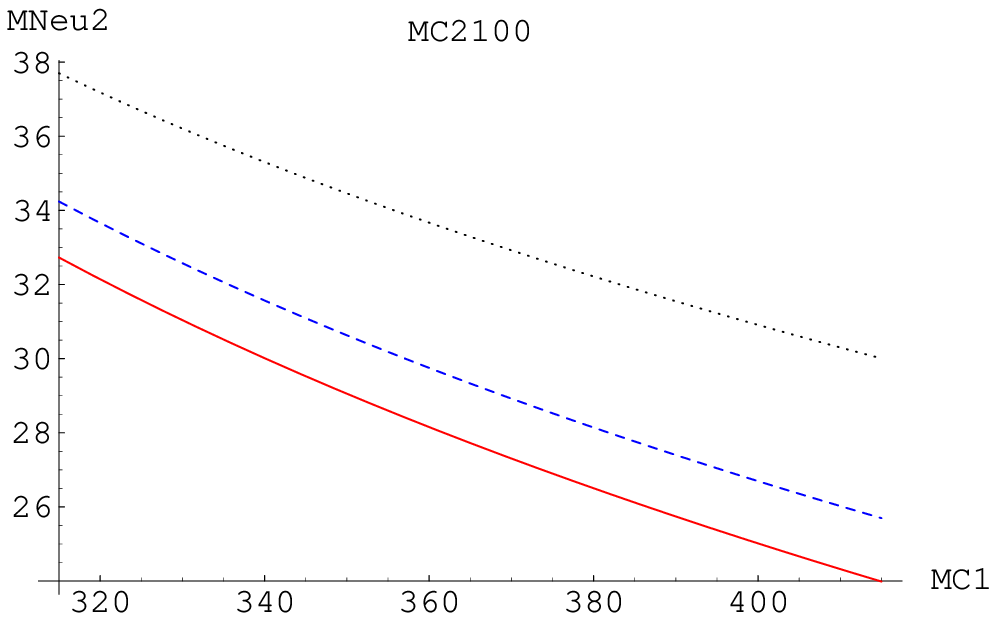,width=0.48\linewidth}
\hfil
\epsfig{file=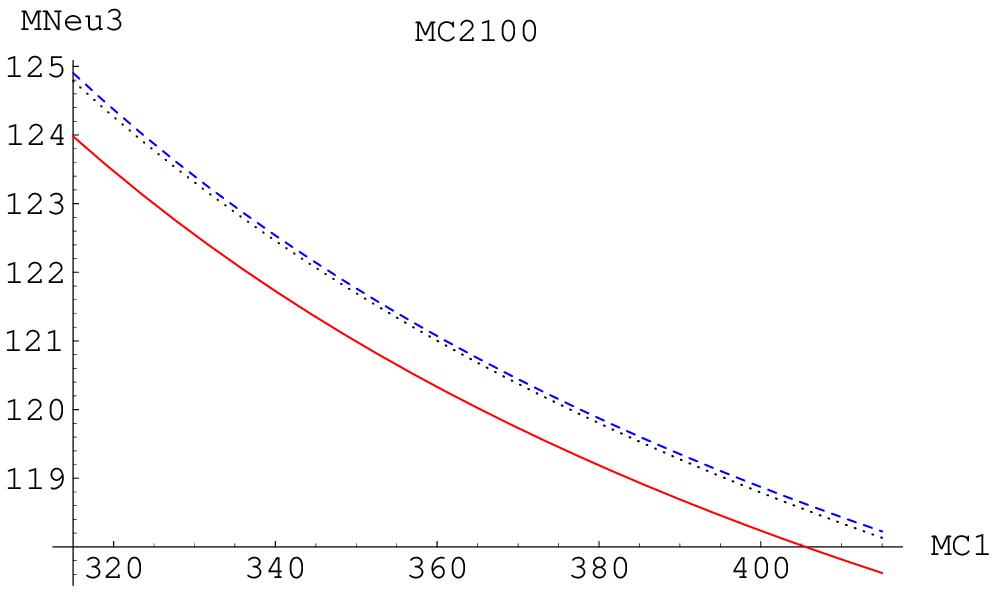,width=0.48\linewidth}
\caption{\footnotesize
Dependence of the calculated neutralino masses (in GeV) 
on the chargino masses $m_{\chrg{1}}$ and  $m_{\chrg{2}}$, 
in Born approximation (dotted, black), including loop corrections with
(s)fermions only (dashed, blue), and with the complete one-loop
contributions (solid, red). The input neutralino mass is chosen as 
$m_{\ntrl{1}}=110$ GeV throughout all diagrams. The plots for 
$m_{\ntrl{4}}$ neutralino would look very much alike the one shown
for all three different values of $m_{\chrg{1}}$.
\label{fig:3DMChaHeavyMNeu}
} 
\end{center}
\end{figure}

Fig.~\ref{fig:3DMChaHeavyMNeu} shows the dependence of the neutralino
masses on the input mass $m_{\chrg{2}}$ of the heavy chargino at three
different values for the mass  $m_{\chrg{1}}$ of the light chargino.
For definiteness we choose the $a_+$ solution of the first line
in~(\ref{eqn:CPInvOnShellRechnung2}) as a representative example
(the other solutions show similar behaviour).
The input mass from the neutralino sector is assumed to be
$m_{\ntrl{1}}=110$ GeV throughout all the subdiagrams of
Fig.~\ref{fig:3DMChaHeavyMNeu}.
Depicted are: the tree-level approximation of the calculated neutralino
masses, their values after including the corrections from the
(s)fermionic loops only, and finally the complete one-loop corrected
masses with all MSSM particles in the virtual states.
For  the heaviest neutralino mass  $m_{\ntrl{4}}$ the shift is small,
not more than 175 MeV throughout the whole scanned parameter space, and
nearly invisible in the graphical illustration. Accordingly, the
relative corrections are less than $0.05$~\% everywhere, and hence we do
not give more than one graph in the figure. 

The impression prima facie of a qualitatively different behaviour
of the masses $m_{\ntrl{2}}$ and $m_{\ntrl{3}}$ in the different
mass regions of $m_{\chrg{1}}$ in Fig. \ref{fig:3DMChaHeavyMNeu} 
can be explained as follows.
Starting from $m_{\chrg{1}}=170$ GeV and going to $m_{\chrg{1}}=135$ GeV
there is a point where the two neutralino-mass curves under
consideration cross each other. At this specific value for
$m_{\chrg{1}}$, the two particles are renumbered reflecting the changed
order of their masses. The different form of this two mass curves for 
$m_{\chrg{1}}=135$ GeV and $m_{\chrg{1}}=100$ GeV stems from the fact
that there is a formal singularity of $M_1$ at a value of 
$m_{\chrg{1}}$ close to  $115$ GeV in 
(\ref{eqn:CPInvOnShellRechnung6}).

The one-loop corrections for the mass $m_{\ntrl{2}}$ 
can reach $20$~\% in the case that the mass splitting between the
two charginos is large. For the second and third neutralino, the typical
size of the loop corrections to the masses amount to about one per cent
of the Born values. 

A synopsis of all diagrams clearly shows that the one-loop contributions
of the gauge and Higgs sector to the neutralino mass shifts are
basically of the same order of magnitude as those resulting from the 
subclass of (s)fermionic loops. 

\clearpage

\subsubsection{Dependence of the neutralino masses on $\tan\beta$}

\begin{figure}[H]
\begin{center}
\vspace{0.5cm}
\psfrag{MC2180}{\footnotesize $m_{\chrg{1}}=180$ GeV}
\psfrag{MDi2}{\tiny $m_{\ntrl{2}}^{\rm Loop}-m_{\ntrl{2}}^{\rm Born}$}
\psfrag{MDi3}{\tiny $m_{\ntrl{3}}^{\rm Loop}-m_{\ntrl{3}}^{\rm Born}$}
\psfrag{MDi4}{\tiny $m_{\ntrl{4}}^{\rm Loop}-m_{\ntrl{4}}^{\rm Born}$}
\psfrag{MNeu2}{\footnotesize $m_{\ntrl{2}}$}
\psfrag{MNeu3}{\footnotesize $m_{\ntrl{3}}$}
\psfrag{MNeu4}{\footnotesize $m_{\ntrl{4}}$}
\psfrag{Tanb}{\footnotesize $\tan\beta$}
\epsfig{file=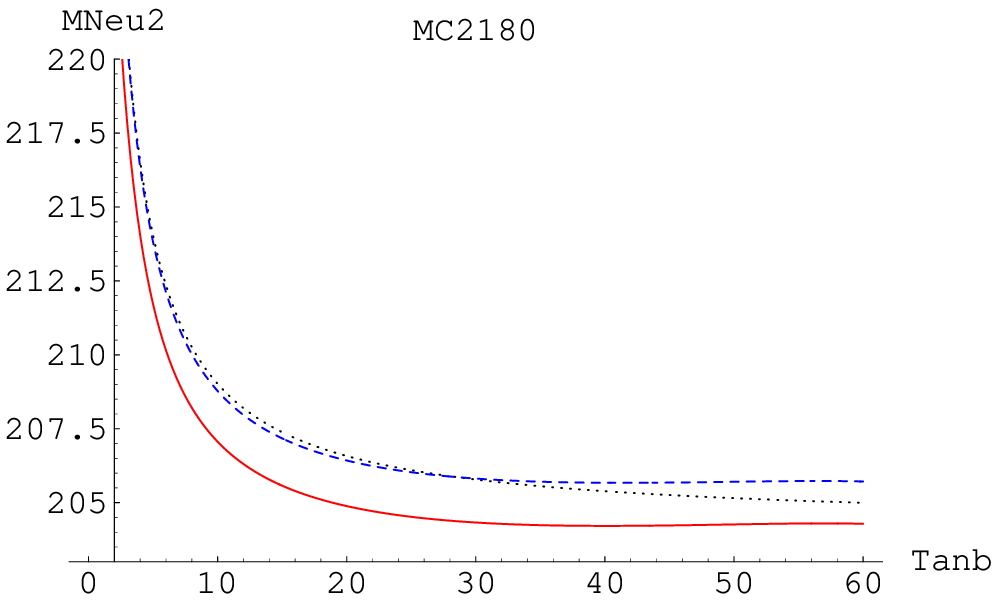,width=0.45\linewidth}
\hfil
\epsfig{file=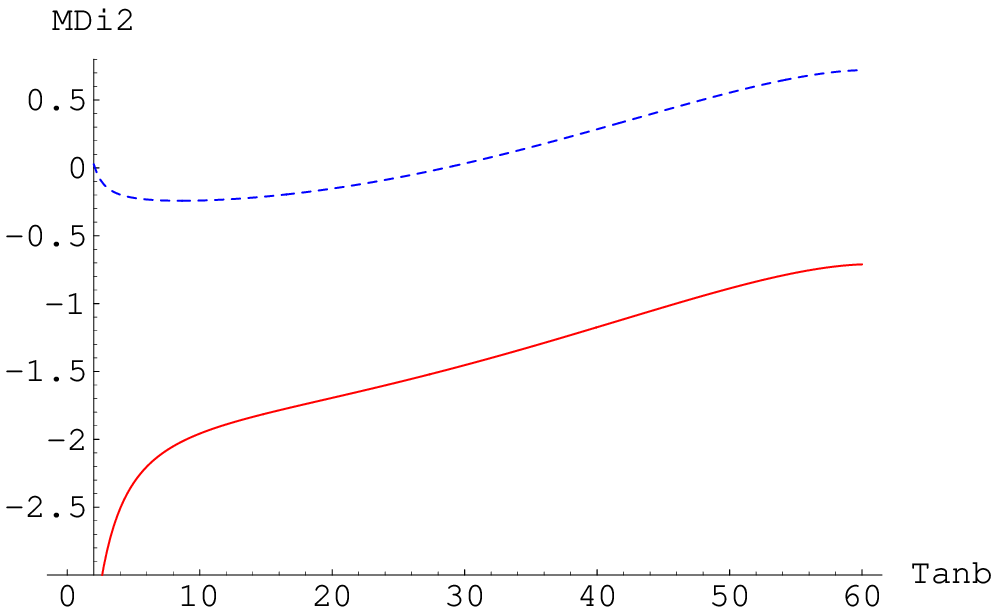,width=0.45\linewidth}
\\[0.2cm]
\epsfig{file=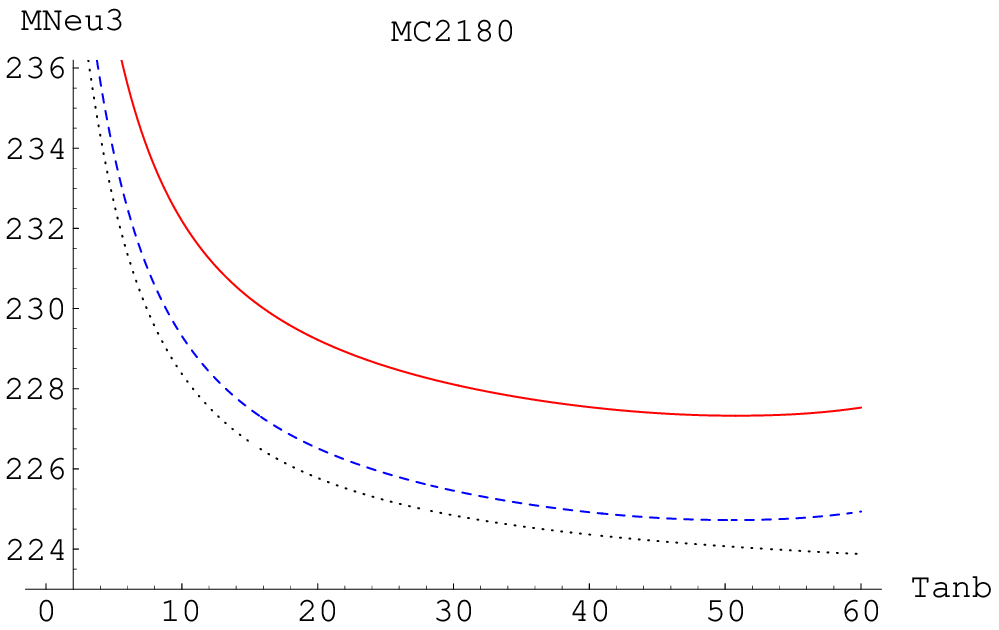,width=0.45\linewidth}
\hfil
\epsfig{file=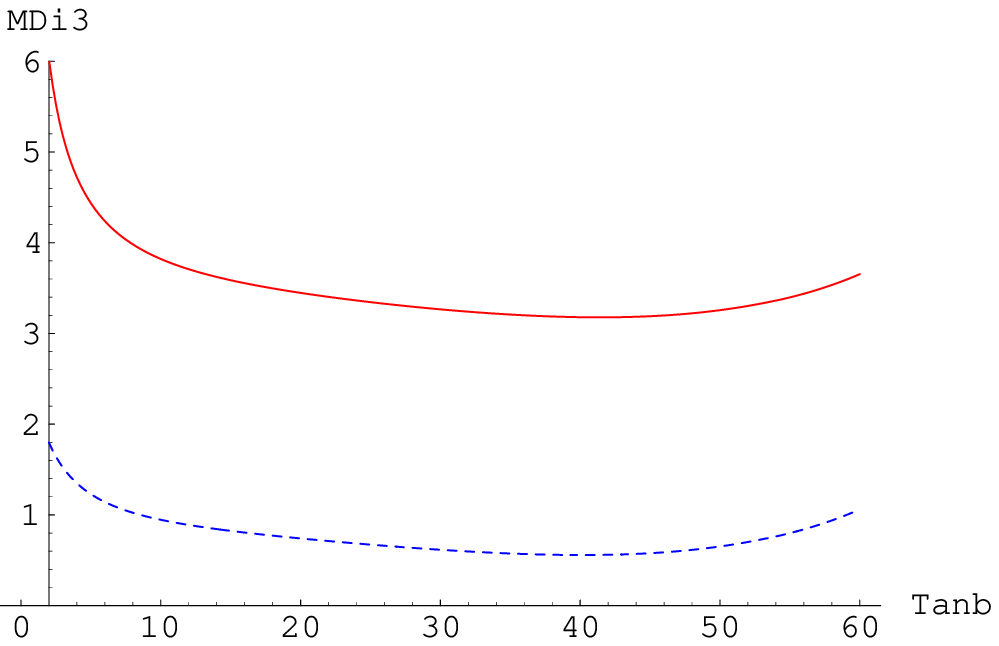,width=0.45\linewidth}
\\[0.2cm]
\epsfig{file=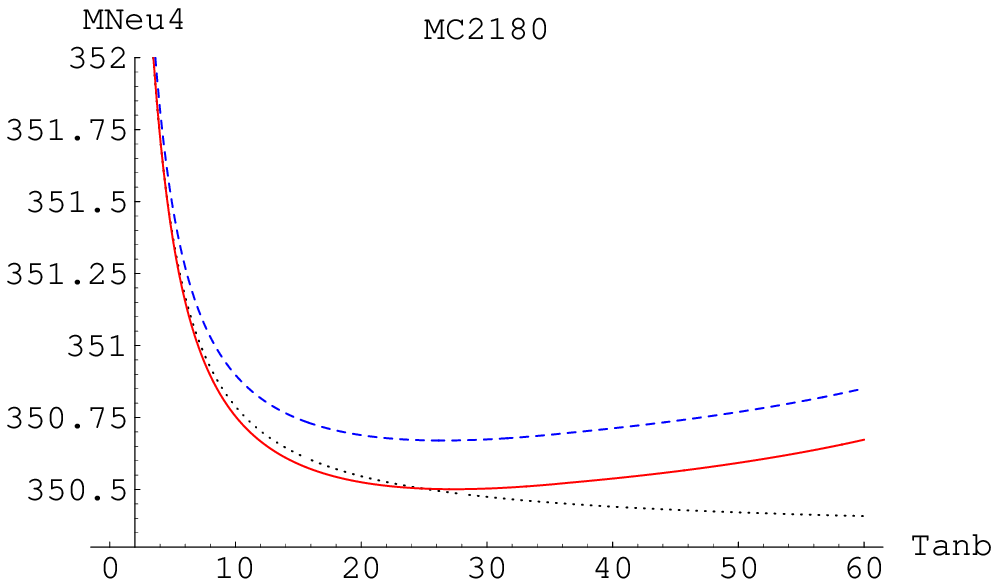,width=0.45\linewidth}
\hfil
\epsfig{file=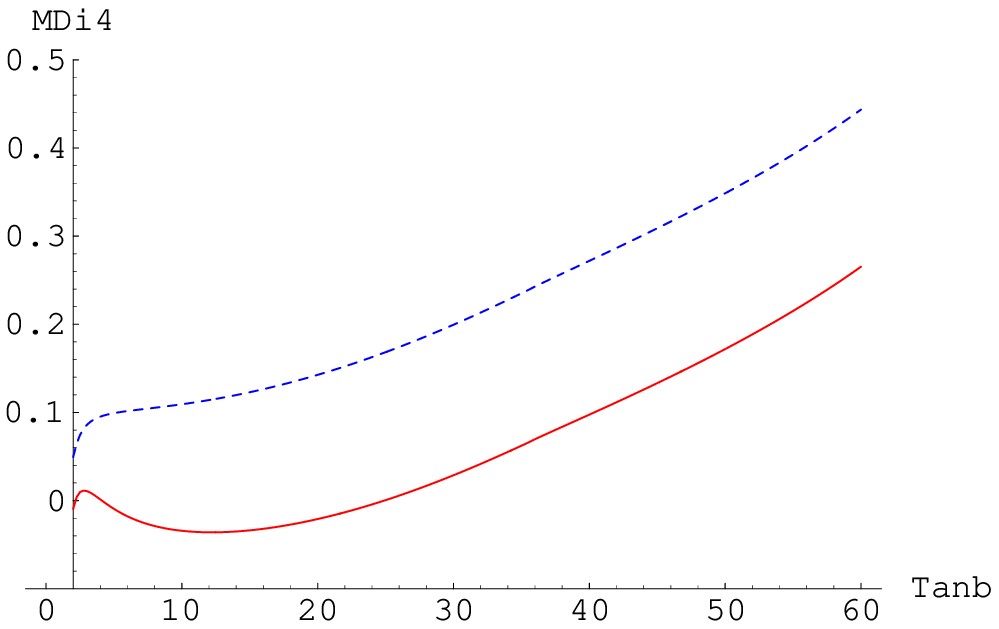,width=0.45\linewidth}
\\[0.2cm]
\caption{\footnotesize
Dependence of the neutralino masses on $\tan \beta$, in Born
approximation (dotted, black), including loop corrections with
(s)fermions only (dashed, blue), and with the complete one-loop
contributions (solid, red). 
The mass of the heavy chargino is set to $m_{\chrg{2}}=350$ GeV
and the input neutralino mass is chosen as $m_{\ntrl{1}}=160$ GeV
throughout all the plots.
Left columns: absolute mass values; right columns: mass shifts (in GeV).
\label{fig:Abb9}
} 
\end{center}
\end{figure}

\begin{figure}[H]
\begin{center}
\vspace{1cm}
\psfrag{MC2100}{\footnotesize $m_{\chrg{1}}=100$ GeV}
\psfrag{MDi2}{\tiny $m_{\ntrl{2}}^{\rm Loop}-m_{\ntrl{2}}^{\rm Born}$}
\psfrag{MDi3}{\tiny $m_{\ntrl{3}}^{\rm Loop}-m_{\ntrl{3}}^{\rm Born}$}
\psfrag{MDi4}{\tiny $m_{\ntrl{4}}^{\rm Loop}-m_{\ntrl{4}}^{\rm Born}$}
\psfrag{MNeu2}{\footnotesize $m_{\ntrl{2}}$}
\psfrag{MNeu3}{\footnotesize $m_{\ntrl{3}}$}
\psfrag{MNeu4}{\footnotesize $m_{\ntrl{4}}$}
\psfrag{Tanb}{\footnotesize $\tan\beta$}
\epsfig{file=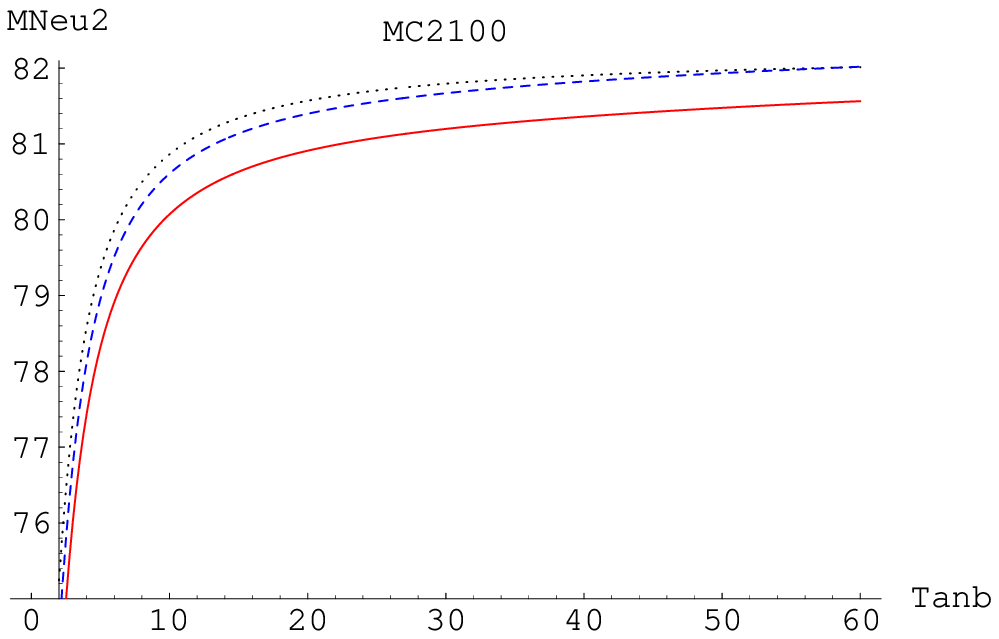,width=0.45\linewidth}
\hfil
\epsfig{file=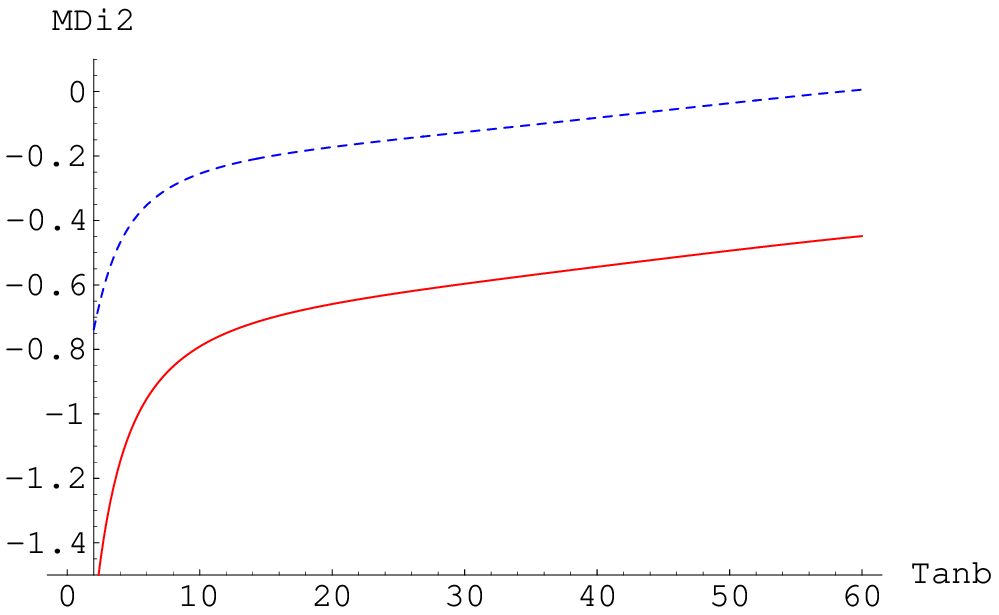,width=0.45\linewidth}
\\[0.2cm]
\epsfig{file=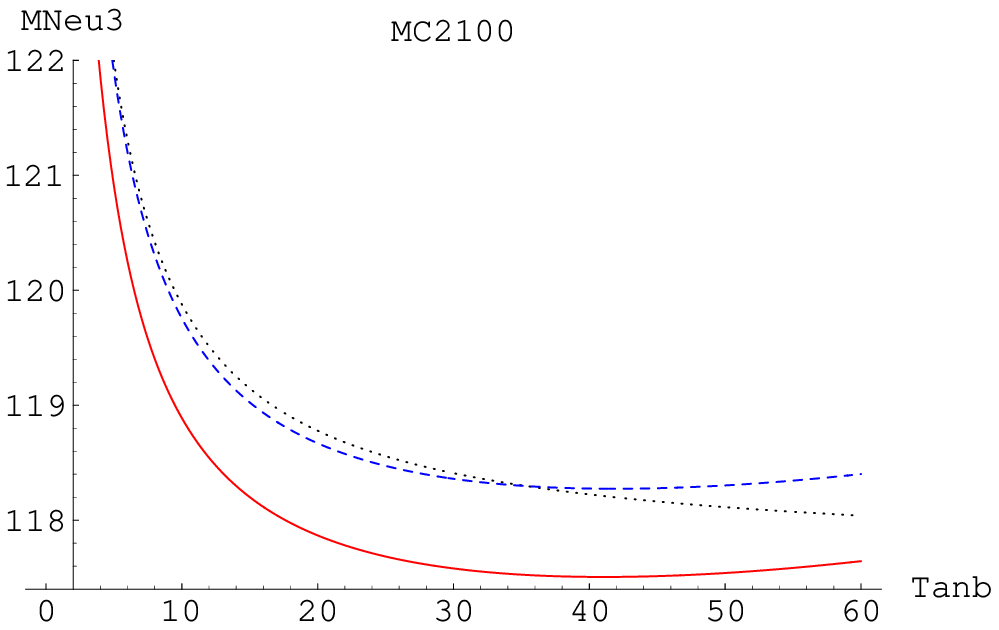,width=0.45\linewidth}
\hfil
\epsfig{file=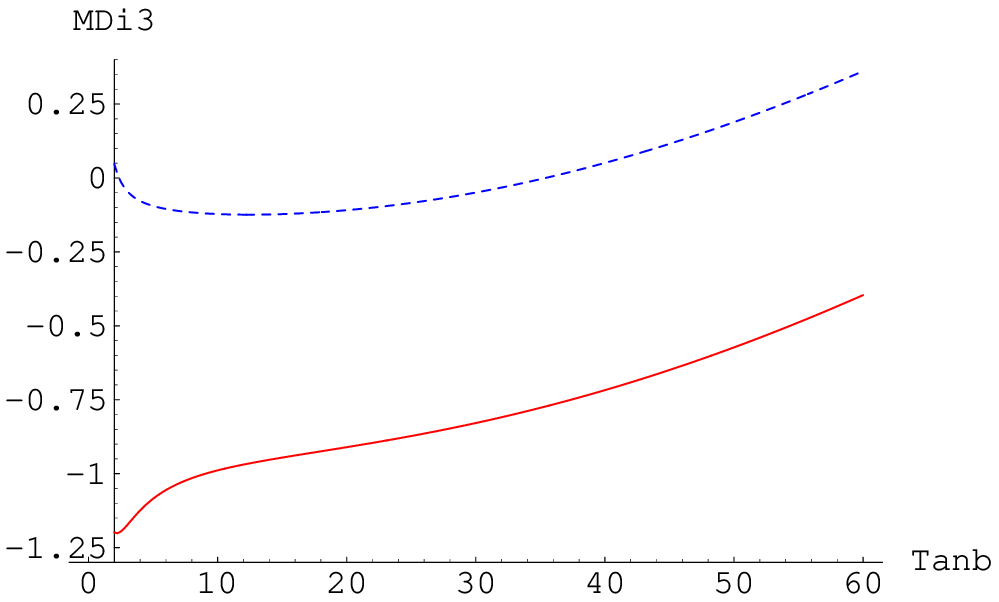,width=0.45\linewidth}
\\[0.2cm]
\epsfig{file=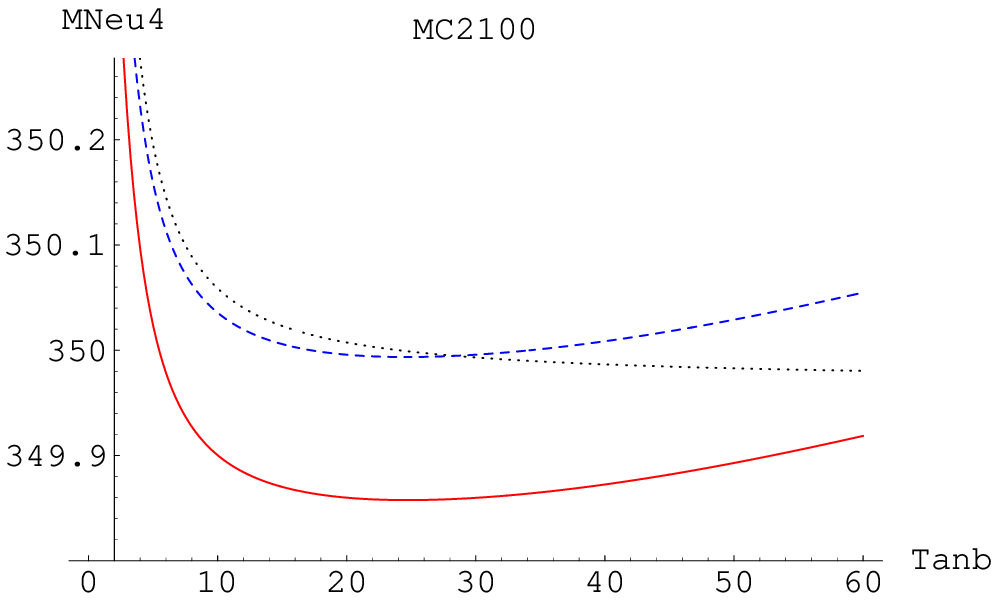,width=0.45\linewidth}
\hfil
\epsfig{file=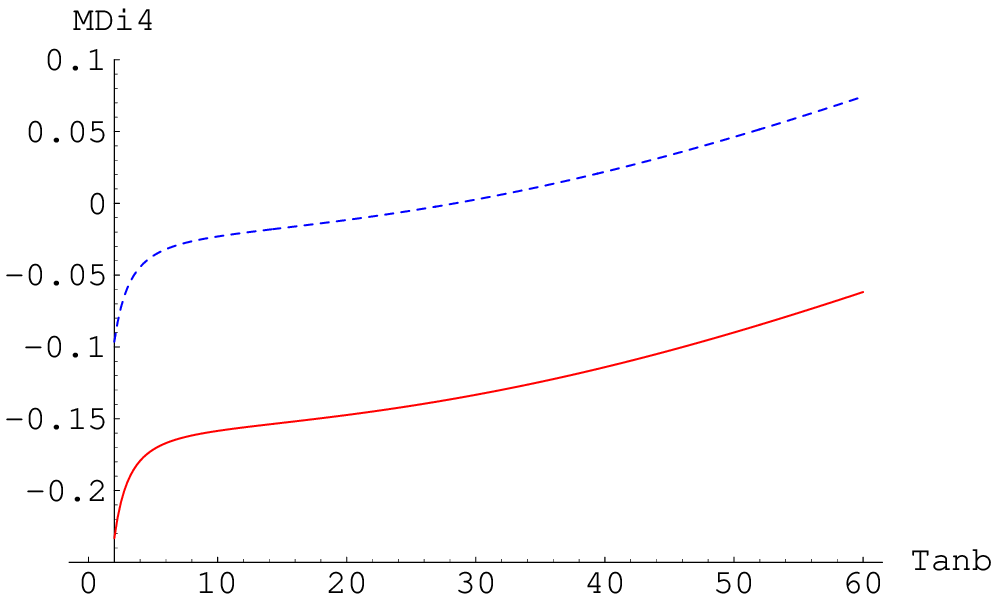,width=0.45\linewidth}
\\[0.2cm]
\caption{\footnotesize
Dependence of the neutralino masses on $\tan \beta$ in Born
approximation (dotted, black), including loop-corrections with
(s)fermions only (dashed, blue) and with the complete one-loop radiative
corrections (solid, red). 
The mass of the heavy chargino is set to $m_{\chrg{2}}=350$ GeV
and the input neutralino mass is chosen as $m_{\ntrl{1}}=160$ GeV
throughout all diagrams. Left columns: 
absolute mass values; right columns: mass shifts (in GeV).
\label{fig:Abb10}
} 
\end{center}
\end{figure}

In Fig. \ref{fig:Abb9} and \ref{fig:Abb10} the $\tan\beta$-dependence
of the calculated neutralino masses is visualized for two different
examples of the light-chargino mass $m_{\chrg{1}}$.
The values for $\tan\beta$ are varied  from 2 to 60.
For the mass of the heavy chargino we choose
$m_{\chrg{2}}=350$ GeV, and the input neutralino mass is set
to $m_{\ntrl{1}}=160$ GeV.

The left columns contain the predicted neutralino masses, again in Born
approximation and at the one-loop level taking into account all MSSM
particles in the loops.
For comparison, the one-loop neutralino masses based on the
approximation with (s)fermionic loops only are also shown.
In the right columns the loop-induced mass shifts (in GeV) are displayed. 

The variation of the one-loop shifts of the neutralino masses over the
whole range of $\tan\beta$ is highest for the case of $\ntrl{2}$. At low
values for $\tan\beta$ the mass correction is about $2$~\% of the Born
value, decreasing steadily with raising $\tan\beta$ to approximately
$0.7$~\%. 

In the parameter space being considered here the mass correction for the
next-heaviest neutralino ranges from $1.5$~\% to $2.5$~\% of the Born
value in the case of $m_{\chrg{1}}=100$ GeV.  
It decreases for heavier charginos, as e.g.\ in the example of
$m_{\chrg{1}}=180$ GeV where it varies between  $0.4$~\% and $0.9$~\%.

The one-loop corrections of the mass of the heaviest neutralino are
small, below $0.1$~\% of the respective Born values throughout the
entire range of $\tan\beta$ under investigation. 

A final remark addresses the option of performing a 
$\overline{\rm MS}$ renormalization of $\tan\beta$, where only
the $\overline{\rm MS}$ UV-singularity of the r.h.s.\ 
in~(\ref{eqn:OtherRCx3}) is defined as the counterterm.
This option has been installed in the version {\it FeynHiggs}1.2 of
the {\it FeynHiggs} code to calculate the neutral MSSM Higgs-boson 
masses and couplings~\cite{FeynHiggs}.
As a comparison of the respectively calculated neutralino masses  
with the neutralino-mass results based on (\ref{eqn:OtherRCx3})
has shown~\cite{Freitas}, the differences  are quite small, at most
about 40 MeV, for the same input values in both 
schemes.\footnote{We thank A. Freitas for the numerical comparison.}

\section{Conclusions}
 
We have presented an on-shell renormalization of the chargino and
neutralino mass spectrum of the MSSM at the one-loop level, based on the
entire set of one-loop diagrams.
An on-shell renormalization scheme has been specified treating all
particle masses as pole masses, with renormalization constants
implemented in a way that allows one to formulate the renormalized
self-energies of the charginos and neutralinos as UV-finite matrices 
which are diagonal for external momenta on-shell.
With the masses of both charginos and of one neutralino as input,
the MSSM parameters $\mu, M_2,M_1$ formally obey the lowest-order
relations to these masses. The masses of the residual three
neutralinos are calculated from the input, yielding mass shifts
up to several GeV as compared to the tree-level approximation.
The numerical investigation shows that the virtual contributions 
beyond those from the subset of diagrams with fermion/sfermion
loops are in general of similar size as the purely (s)fermionic
contributions. A proper treatment will therefore become necessary for 
precision studies within the MSSM at future colliders.

\section*{Acknowledgement}

This work was supported in part by the DFG Forschergruppe 
``Quantenfeldtheorie, Computeralgebra und Monte-Carlo-Simulation''
and by the European Community's Human Potential Programme under contract
HPRN-CT-2000-00149 ``Physics at Colliders''. 
The calculations have been performed using the QCM cluster of the DFG
Forschergruppe.  
We want to thank H. Eberl, A. Freitas, J. Guasch, and W. Majerotto for
useful discussions and for various numerical comparisons.

\end{document}